\def\beq#1{\begin{equation}\label{#1}}
\def\eeq{\end{equation}}
\def\beqa#1{\begin{eqnarray}\label{#1}}
\def\eeqa{\end{eqnarray}}

\def\fun#1#2{\lower3.6pt\vbox{\baselineskip0pt\lineskip.9pt
        \ialign{$\mathsurround=0pt#1\hfill##\hfil$\crcr#2\crcr\sim\crcr}}}

\def\xi{{{\bf x}^b}}

\documentclass
[twocolumn,aps,showpacs,showkeys,nofootinbib,superscriptaddress]{revtex4}
\usepackage{epsfig}

\newcommand{\be}{\begin{equation}}
\newcommand{\ee}{\end{equation}}
\newcommand{\ba}{\begin{eqnarray}}
\newcommand{\ea}{\end{eqnarray}}

\begin{document}
\input{epsf.sty}

\title{Revisit of constraints on holographic dark energy:
SNLS3 dataset with the effects of time-varying $\beta$ and different light-curve fitters}

\author{Shuang Wang}
\affiliation{Department of Physics, College of Sciences, Northeastern University, Shenyang 110004, China}

\author{Jia-Jia Geng}
\affiliation{Department of Physics, College of Sciences, Northeastern University, Shenyang 110004, China}

\author{Yi-Liang Hu}
\affiliation{Department of Physics, College of Sciences, Northeastern University, Shenyang 110004, China}

\author{Xin Zhang\footnote{Corresponding author.}}
\email{zhangxin@mail.neu.edu.cn}
\affiliation{Department of Physics, College of Sciences, Northeastern University, Shenyang 110004, China}
\affiliation{Center for High Energy Physics, Peking University, Beijing 100080, China}

\begin{abstract}

Previous studies have shown that
for the Supernova Legacy Survey three-year (SNLS3) data
there is strong evidence for the redshift-evolution of color-luminosity parameter $\beta$ of type Ia supernovae (SN Ia).
In this paper, we explore the effects of varying $\beta$ on the cosmological constraints of holographic dark energy (HDE) model.
In addition to the SNLS3 data, we also use Planck distance prior data of cosmic microwave background (CMB),
as well as galaxy clustering (GC) data extracted from
Sloan Digital Sky Survey (SDSS) data release 7 and Baryon Oscillation Spectroscopic Survey (BOSS).
We find that, for the both cases of using SN data alone and using SN+CMB+GC data,
involving an additional parameter of $\beta$ can reduce $\chi^2$ by $\sim$ 36;
this shows that $\beta$ deviates from a constant at 6$\sigma$ confidence levels.
Adopting SN+CMB+GC data,
we find that compared to the constant $\beta$ case,
varying $\beta$ yields a larger fractional matter density $\Omega_{m0}$ and a smaller reduced Hubble constant $h$;
moreover, varying $\beta$ significantly increases the value of HDE model parameter $c$,
leading to $c\approx 0.8$, consistent with the constraint results obtained before Planck.
These results indicate that the evolution of $\beta$ should be taken into account seriously in the cosmological fits.
In addition, we find that relative to the differences between the constant $\beta$ and varying $\beta(z)$ cases,
the effects of different light-curve fitters on parameter estimation are very small.

\end{abstract}

\pacs{98.80.-k, 98.80.Es, 95.36.+x}

\keywords{Cosmological constraints, holographic dark energy, type Ia supernova, time-varying $\beta$}

\maketitle

\section{Introduction}

Since its discovery,
cosmic acceleration has become one of the most important research fields in modern cosmology
\cite{Riess98,spergel03,Tegmark04,Komatsu09,Percival10,Drinkwater10,Riess11}.
Cosmic acceleration may be due to an unknown energy component
(i.e., dark energy (DE) \cite{quint,phantom,k,Chaplygin,ngcg,tachyonic,hessence,YMC,hscalar,cq,others1,others2,WangTegmark05,others3}),
or a modification of general relativity (i.e., modified gravity (MG) \cite{SH,PR,DGP,GB,Galileon,FR,FT,FRT}).
For recent reviews, see, e.g., \cite{CST,FTH,Linder,CK,Uzan,Tsujikawa,NO,LLWW,CFPS,YWBook}.

One of the most powerful probes of DE is the use of type Ia supernovae (SN Ia)
\cite{Union,Constitution,Union2,Union2.1}.
In 2010, the Supernova Legacy Survey (SNLS) group released their three-years data, i.e. SNLS3 dataset \cite{SNLS3}.
Soon after, Conley et al. presented SN-only cosmological results
by combining the SNLS3 SNe with various low- to mid-$z$ samples \cite{SNLS3Conley},
and Sullivan et al. presented the joint cosmological constraints
by combining the SNLS3 dataset with other cosmological observations \cite{SNLS3Sullivan}.
Depending on different light-curve fitters, Conley et al. \cite{SNLS3Conley} presented three SN data sets:
``SALT2'', which consists of 473 SNe Ia;
``SiFTO'', which consists of 468 SNe Ia;
and ``Combined'', which consists of 472 SNe Ia.
Unlike other SN groups, the SNLS team treated two important quantities,
stretch-luminosity parameter $\alpha$ and color-luminosity parameter $\beta$ of SNe Ia,
as free model parameters.

Currently, a critical challenge is the control of the systematic uncertainties of SNe Ia.
One of the most important factors is the effect of potential SN evolution,
i.e. the possibility for the redshift evolution of $\alpha$ and $\beta$.
Current studies show that $\alpha$ is still consistent with a constant,
but the hints of evolution of $\beta$ have been found in \cite{Astier06,Kessler09,Marriner11,Scolnic1,Scolnic2}.
In \cite{Mohlabeng}, Mohlabeng and Ralston studied the case of Union2.1 sample using $\beta(z) = \beta_0 + \beta_1 z$,
and found that $\beta$ deviates from a constant at 7$\sigma$ confidence levels (CL).
In \cite{WangWang}, using the SNLS3 data, Wang and Wang found that
$\beta$ increases significantly with $z$ at the 6$\sigma$ CL when systematic uncertainties are taken into account;
moreover, they proved that this conclusion is insensitive to the lightcurve fitter used to derive the SNLS3 sample,
or the functional form of $\beta(z)$ \cite{WangWang}.
Therefore, the evolution of $\beta$ should be taken into account seriously in the cosmology fits.

It is clear that the evolution of $\beta$ will have significant effects.
In \cite{WangNew},
using the $\Lambda$-cold-dark-matter ($\Lambda$CDM) model, the $w$CDM model, and the Chevallier-Polarski-Linder (CPL) model,
Wang, Li, and Zhang showed that adding a parameter of $\beta$ could significantly improve the fitting results;
in addition, considering the evolution of $\beta$ is helpful in reducing the tension between SN and other cosmological observations.
It should be pointed out that all the models considered in \cite{WangNew} are very simple.
To further study the issue of varying $\beta$,
some more specific DE models need to be taken into account.
In this paper,
we study the effects of a time-varying $\beta$ on parameter estimation in the holographic dark energy (HDE) model \cite{Li1}.
The HDE is a physically plausible DE candidate based on the holographic principle \cite{Holography};
it has been widely studied both theoretically \cite{HDEworks} and observationally \cite{HDEObserv}.

We first briefly review the theoretical framework of the HDE model.
In \cite{Cohen}, Cohen {\it et al.} suggested that quantum zero-point energy of a system with size $L$
should not exceed the mass of a black hole with the same size,
i.e., $L^3 k_{\rm max}^4\leq L M^2_{\rm Pl}$
(here $M_{\rm Pl}$ is the reduced Planck mass, and $k_{\rm max}$ is the ultraviolet (UV) cutoff of the system).
Therefore, the UV cutoff of a system is related to its infrared (IR) cutoff.
When we consider the whole universe,
the vacuum energy related to this holographic principle can be
viewed as dark energy, and the corresponding energy density becomes
\begin{equation}\label{eq:rhohde}
 \rho_{\rm de}=3c^2M^2_{\rm Pl}L^{-2},
\end{equation}
where $c$ is a dimensionless model parameter that modulates the DE density \cite{Li1}.
In \cite{Li1}, Li suggested that the
IR length-scale cutoff should be chosen as the size of the future event horizon of the universe, $R_{eh}(t)=a(t)\int_t^{+\infty}dt'/a(t')$.
More generically, when we also consider the spatial curvature in a universe, the IR length cut-off $L$ takes the form
\begin{equation}\label{eq:Rh}
 L=ar(t),
\end{equation}
where
\begin{equation}
r(t)={1\over \sqrt{k}}{\rm sinn}\left(\sqrt{k}\int_t^{+\infty}{dt'\over a(t')}\right),
\end{equation}
with ${\rm sinn}(x)=\sin(x)$, $x$, and $\sinh(x)$ for $k>0$, $k=0$, and $k<0$, respectively.
This leads to the following equation of state (EOS) of DE,
\begin{equation}\label{eq:hdew}
 w_{\rm de}(z)=-\frac{1}{3}-\frac{2}{3}\sqrt{{\Omega_{\rm de}(z)\over c^2}+\Omega_k(z)},
\end{equation}
which can yield an accelerated universe.
In Eq.~(\ref{eq:hdew}),
the function $\Omega_{\rm de}(z)$ is determined by the following coupled differential equation system \cite{LWLZ},
\begin{equation}
\label{eq:deq1}{1\over E}{dE \over dz} =-{\Omega_{de}\over 1+z}
\left({\Omega_k-\Omega_{r}-3\over2\Omega_{de}}+{1\over2} +\sqrt{{\Omega_{de}\over c^2}+\Omega_k} \right),
\end{equation}
\begin{equation}
\label{eq:deq2} {d\Omega_{de}\over dz}= -{2\Omega_{de}(1-\Omega_{de})\over 1+z}
\left(\sqrt{{\Omega_{de}\over c^2}+\Omega_k}+{1\over2}-{\Omega_k-\Omega_{r}\over 2(1-\Omega_{de})}\right),
\end{equation}
where $E(z)\equiv H(z)/H_0$ is the dimensionless Hubble expansion rate,
$H_0=100h$ km s$^{-1}$ Mpc$^{-1}$ is the Hubble constant,
$\Omega_k(z)=\Omega_{k0}(1+z)^2/E(z)^2$, and $\Omega_{r}(z)=\Omega_{r0}(1+z)^4/E(z)^2$.
In addition, $\Omega_{r0}=\Omega_{m0} / (1+z_{\rm eq})$,
$z_{\rm eq}=2.5\times 10^4 \Omega_{m0} h^2 (T_{\rm cmb}/2.7\,{\rm K})^{-4}$
(here we take $T_{\rm cmb}=2.7255\,{\rm K}$).
The initial conditions are $E(0)=1$
and $\Omega_{de}(0)=1-\Omega_{m0}-\Omega_{k0}-\Omega_{r0}$.
By numerically solving Eqs.~(\ref{eq:deq1}) and (\ref{eq:deq2}),
we can obtain the evolution of $E(z)$, which can be used to calculate all the observational quantities appearing in Sec.~II.

In this paper, we explore the effects of varying $\beta$ on the SNLS3 constraints of the HDE model.
In addition to the SNLS3 data,
we also use the Planck distance prior data \cite{WangWangCMB},
as well as the latest galaxy clustering (GC) data
extracted from SDSS DR7 \cite{ChuangWang12} and BOSS \cite{Chuang13}.
In addition, we also study the effects of different light-curve fitters on parameter estimation.

We describe our method in Sec.~II, present our results in Sec.~III, and conclude in Sec.~IV.
In this paper, we assume today's scale factor $a_{0}=1$, thus the redshift $z=a^{-1}-1$.
The subscript ``0'' always indicates the present value of the corresponding quantity, and the
natural units are used.

\section{Method}
\label{sec:method}

In this section, we will introduce how to include the SNLS3 data into the $\chi^2$ analysis.

The comoving distance to an object at redshift $z$ is given by
\be
\label{eq:r(z)}
 r(z)=H_0^{-1}\, |\Omega_{k0}|^{-1/2} {\rm sinn}[|\Omega_{k0}|^{1/2}\, \Gamma(z)],
\ee
where $\Gamma(z)=\int_0^z\frac{dz'}{E(z')}$,
and ${\rm sinn}(x)=\sin(x)$, $x$, $\sinh(x)$ for $\Omega_{k0}<0$, $\Omega_{k0}=0$, and $\Omega_{k0}>0$ respectively.

SN Ia data give measurements of the luminosity distance $d_L(z)$
through that of the distance modulus of each SN:
\be
\label{eq:m-M}
\mu_0 \equiv m-M= 5 \log\left[\frac{d_L(z)}{\mathrm{Mpc}}\right]+25,
\ee
where $m$ and $M$ represent the apparent and absolute magnitude of an SN.
Moreover, the luminosity distance $d_L(z)=(1+z)\, r(z)$.

Here we use the SNLS3 data set.
As mentioned above, based on different light-curve fitters,
three SN sets of SNLS3 are given, including ``SALT2'', ``SiFTO'', and ``Combined''.
To perform a comparative study, all these three sets will be used in this paper.

In \cite{WangWang},
by considering three functional forms (linear case, quadratic case, and step function case),
Wang and Wang showed that the evolutions of $\alpha$ and $\beta$
are insensitive to functional form of $\alpha$ and $\beta$.
So in this paper,
we just adopt a constant $\alpha$ and a linear $\beta(z) = \beta_{0} + \beta_{1} z$.
Now, the predicted magnitude of an SN becomes
\be
m_{\rm mod}=5 \log_{10}{\cal D}_L(z|\mbox{\bf p})
- \alpha (s-1) +\beta(z) {\cal C} + {\cal M},
\ee
where ${\cal D}_L(z|\mbox{\bf p})$ is the luminosity distance
multiplied by $H_0$
for a given set of cosmological parameters $\{ {\bf p} \}$,
$s$ is the stretch measure of the SN light curve shape,
and ${\cal C}$ is the color measure for the SN.
${\cal M}$ is a nuisance parameter representing some combination
of the absolute magnitude of a fiducial SN, $M$, and the Hubble constant, $H_0$.
It must be emphasized that,
to include host-galaxy information in the cosmological fits,
Conley et al. \cite{SNLS3Conley} split the SNLS3 sample based on host-galaxy stellar mass at $10^{10} M_{\odot}$,
and ${\cal M}$ is allowed to be different for the two samples.
Therefore, unlike other SN samples, there are two values of ${\cal M}$, ${\cal M}_1$ and ${\cal M}_2$, for the SNLS3 data
(for more details, see Sections $3.2$ and $5.8$ of \cite{SNLS3Conley}).
Moreover, Conley et al. removed ${\cal M}_1$ and ${\cal M}_2$ from cosmological fits by analytically marginalizing over them
(for more details, see Appendix C of \cite{SNLS3Conley},
as well as the the public code, which is available at https://tspace.library.utoronto.ca/handle/1807/24512).
In this paper, we just follow the recipe of \cite{SNLS3Conley}.

Since the time dilation part of the observed luminosity distance depends
on the total redshift $z_{\rm hel}$ (special relativistic plus cosmological),
we have
\be
{\cal D}_L(z|\mbox{\bf s}) = c^{-1}H_0 (1+z_{\rm hel}) r(z|\mbox{\bf s}),
\ee
where $z$ and $z_{\rm hel}$ are the CMB restframe and heliocentric redshifts of the SN.

For a set of $N$ SNe with correlated errors, the $\chi^2$ function is \cite{SNLS3Conley}
\be
\label{eq:chi2_SN}
\chi^2_{SN}=\Delta \mbox{\bf m}^T \cdot \mbox{\bf C}^{-1} \cdot \Delta\mbox{\bf m}
\ee
where $\Delta m \equiv m_B-m_{\rm mod}$ is a vector with $N$ components,
$m_B$ is the rest-frame peak B-band magnitude of the SN,
and $\mbox{\bf C}$ is the $N\times N$ covariance matrix of the SN.

The total covariance matrix is \cite{SNLS3Conley}
\be
\mbox{\bf C}=\mbox{\bf D}_{\rm stat}+\mbox{\bf C}_{\rm stat}
+\mbox{\bf C}_{\rm sys},
\ee
with the diagonal part of the statistical uncertainty given by \cite{SNLS3Conley}
\ba
\mbox{\bf D}_{{\rm stat},ii}&=&\sigma^2_{m_B,i}+\sigma^2_{\rm int}
+ \sigma^2_{\rm lensing}+ \sigma^2_{{\rm host}\,{\rm correction}} \nonumber\\
&& + \left[\frac{5(1+z_i)}{z_i(1+z_i/2)\ln 10}\right]^2 \sigma^2_{z,i} \nonumber\\
&& +\alpha^2 \sigma^2_{s,i}+\beta(z_i)^2 \sigma^2_{{\cal C},i} \nonumber\\
&& + 2 \alpha C_{m_B s,i} - 2 \beta(z_i) C_{m_B {\cal C},i} \nonumber\\
&& -2\alpha \beta(z_i) C_{s {\cal C},i},
\ea
where $C_{m_B s,i}$, $C_{m_B {\cal C},i}$, and $C_{s {\cal C},i}$
are the covariances between $m_B$, $s$, and ${\cal C}$ for the $i$-th SN,
$\beta_i=\beta(z_i)$ are the values of $\beta$ for the $i$-th SN.
Note also that $\sigma^2_{z,i}$ includes a peculiar velocity residual of 0.0005
(i.e., 150$\,$km/s) added in quadrature \cite{SNLS3Conley}.
Following \cite{SNLS3Conley}, here we fix the intrinsic scatter $\sigma_{int}$ to ensure that $\chi^2/dof=1$.
Varying $\sigma_{int}$ could have a significant impact on parameter estimation, see \cite{Kim2011} for details.

We define $\mbox{\bf V} \equiv \mbox{\bf C}_{\rm stat} + \mbox{\bf C}_{\rm sys}$,
where $\mbox{\bf C}_{\rm stat}$ and $\mbox{\bf C}_{\rm sys}$
are the statistical and systematic covariance matrices, respectively.
After treating $\beta$ as a function of $z$,
$\mbox{\bf V}$ is given in the form,
\ba
\mbox{\bf V}_{ij}&=&V_{0,ij}+\alpha^2 V_{a,ij} + \beta_i\beta_j V_{b,ij} \nonumber\\
&& +\alpha V_{0a,ij} +\alpha V_{0a,ji} \nonumber\\
&& -\beta_j V_{0b,ij} -\beta_i V_{0b,ji} \nonumber\\
&& -\alpha \beta_j V_{ab,ij} - \alpha \beta_i V_{ab,ji}.
\ea
It must be stressed that, while $V_0$, $V_{a}$, $V_{b}$, and $V_{0a}$
are the same as the ``normal'' covariance matrices
given by the SNLS data archive, $V_{0b}$, and $V_{ab}$ are {\it not} the same as the ones given there.
This is because the original matrices of SNLS3 are produced by assuming that $\beta$ is constant.
We have used the $V_{0b}$ and $V_{ab}$ matrices for the ``Combined'' set
that are applicable when varying $\beta(z)$ (A.~Conley, private communication, 2013).

In addition, to break the degeneracy between various model parameters,
we also use the Planck distance prior data \cite{WangWangCMB},
as well as the latest galaxy clustering (GC) data
extracted from SDSS DR7 \cite{ChuangWang12} and BOSS \cite{Chuang13}.
For details on including Planck and GC data into the $\chi^2$ analysis, see Ref. \cite{WangNew}.
Thus, the total $\chi^2$ function is
\be
\chi^2=\chi^2_{SN}+\chi^2_{CMB}+\chi^2_{GC}.
\ee

\section{Results}

We perform a Markov Chain Monte Carlo (MCMC) likelihood analysis \cite{COSMOMC}
to obtain ${\cal O}$($10^6$) samples for each set of results presented in this paper.
We assume flat priors for all the parameters, and allow ranges of the parameters wide enough
such that further increasing the allowed ranges has no impact on the results.
The chains typically have worst e-values
(the variance(mean)/mean(variance) of 1/2 chains) much smaller than 0.01, indicating convergence.

In the following section, we will discuss the effects of varying $\beta$ and different light-curve fitters
on the SNLS3 constraints on the HDE model, respectively.

\subsection{The effects of varying $\beta$}
\label{sec:varbeta}

In this subsection, we discuss the effects of varying $\beta$.
As mentioned previously, to explore the evolution of $\beta$,
we study the case of constant $\alpha$ and linear $\beta(z) = \beta_{0} + \beta_{1} z$;
for comparison, the case of constant $\alpha$ and constant $\beta$ is also taken into account.
For simplicity, here we only use the SN data from the ``Combined'' set.

\begin{itemize}
 \item SN-only cases
\end{itemize}

Firstly, we discuss the results given by the SN data alone.
Notice that the Hubble constant $h$ has been marginalized during the $\chi^2$ fitting process of SNe Ia,
so for this case, we only need to consider six free parameters, including $\alpha$, $\beta_0$, $\beta_1$, $\Omega_{m0}$, $c$, and $\Omega_{k0}$.

In Table \ref{table1}, we list the fitting results for various constant $\beta$ and linear $\beta(z)$ cases, where only the SNLS3 SNe data are used.
The most obvious feature of this table is that a varying $\beta$ can significantly improve the fitting results of HDE model:
adding a parameter of $\beta$ can reduce the best-fit values of $\chi^2$ by $\sim$ 36.
Based on the Wilk's theorem, 36 units of $\chi^2$ is equivalent to a Gaussian fluctuation of 6$\sigma$.
This means that for HDE model, the result of $\beta_1=0$ is ruled out at 6$\sigma$ CL.
This result is consistent with the cases of the $\Lambda$CDM, the $w$CDM, and the CPL models \cite{WangNew}.
In addition, we find that for both the constant $\beta$ and the linear $\beta(z)$ cases,
using the SNe data alone will lead to unreasonable results of $\Omega_{m0}$, $c$, and $\Omega_{k0}$,
inconsistent with the constraint results given by previous studies \cite{HDECon1,HDECon2,HDECon3,PlanckHDE}.
This implies that using SNe data alone cannot constrain the cosmological parameters well.

\begin{table}
\caption{\textrm{A comparison for the fitting results of constant $\beta$ and linear $\beta(z)$ cases.
Only the SN(Combined) data are used in the analysis.}}
\label{table1}
\begin{tabular}{|c|c|c|}
  \hline
  \textrm{Parameter} & constant $\beta$ case & linear $\beta(z)$ case \\
  \hline
  $\alpha$ & $1.433^{+0.100}_{-0.108}$ & $1.417^{+0.099}_{-0.100}$ \\

  $\beta_0$ & $3.262^{+0.105}_{-0.108}$ & $1.429^{+0.358}_{-0.383}$ \\

  $\beta_1$ & \textrm{N/A} & $5.142^{+1.069}_{-0.998}$ \\

  $\Omega_{m0}$ & $0.112^{+0.056}_{-0.056}$ & $0.090^{+0.052}_{-0.054}$ \\

  $c$ & $1.252^{+0.414}_{-0.434}$ & $1.262^{+0.737}_{-0.533}$ \\

  $\Omega_{k0}$ & $0.090^{+0.247}_{-0.238}$ & $0.343^{+0.199}_{-0.209}$ \\
    \hline
  $\chi^2_{min}$ & $419.579$ & $383.560$ \\
  \hline
\end{tabular}
\end{table}

In addition, we also calculate the best-fit values of ${\cal M}_1$ and ${\cal M}_2$ for the SNe-only cases.
For the constant $\beta$ case,
we get ${\cal M}_1 = 0.00206$ and ${\cal M}_2 = 0.01789$;
for the linear $\beta(z)$ case,
we obtain ${\cal M}_1 = -0.00032$ and ${\cal M}_2 = -0.00559$.
We can see that adding a parameter $\beta_1$ will not significantly change the values of ${\cal M}_1$ and ${\cal M}_2$.
This shows that ${\cal M}$ has no significant effects on the conclusion of $\beta$'s evolution.

\begin{figure}
\includegraphics[scale=0.25, angle=0]{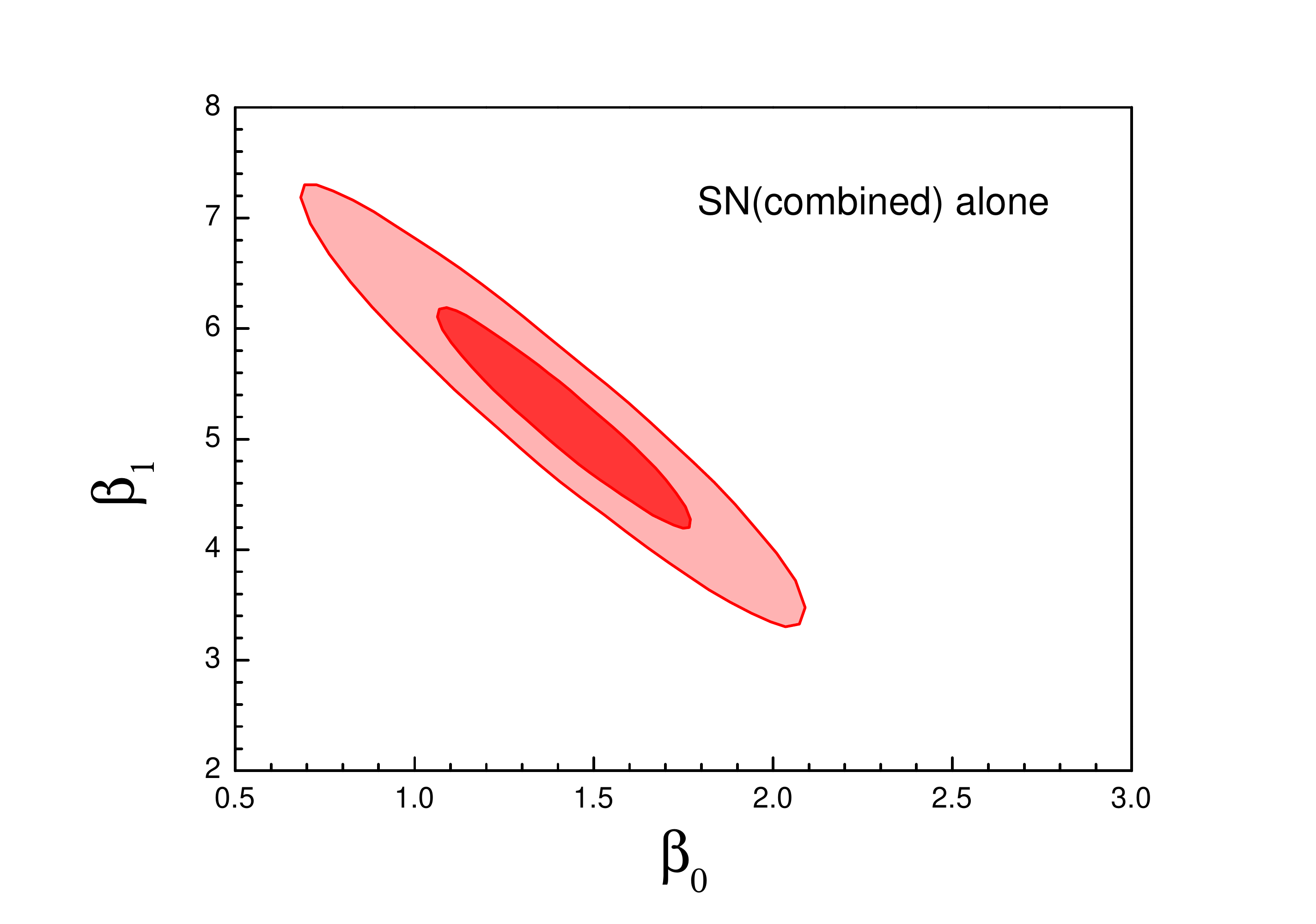}
\includegraphics[scale=0.25, angle=0]{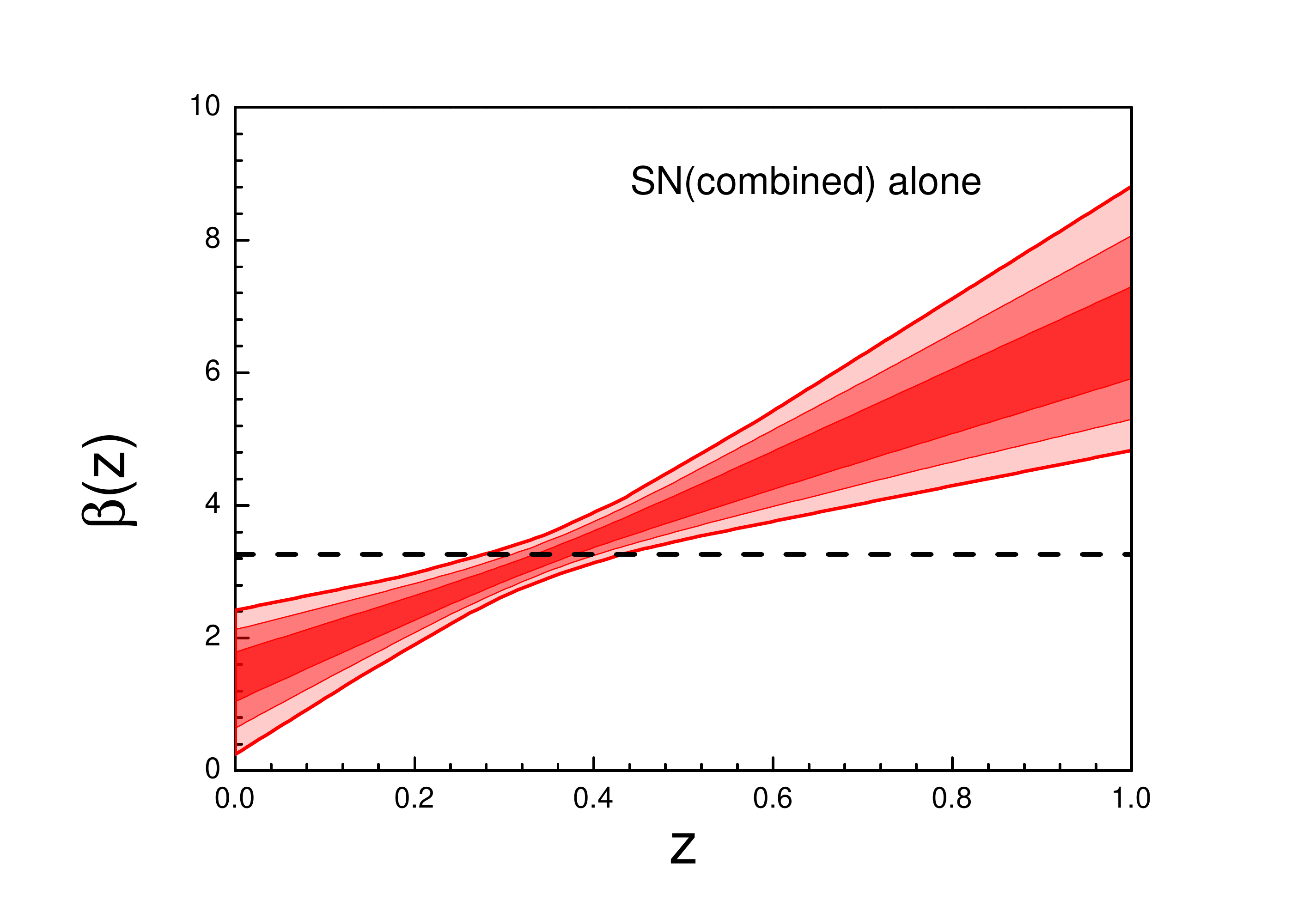}
\caption{\label{fig1}\footnotesize%
The joint $68\%$ and $95\%$ confidence contours for $\{\beta_{0},\beta_{1}\}$ (top panel)
and the $68\%$, $95\%$, and $97\%$ confidence constraints for $\beta(z)$ (bottom panel),
given by the SN(Combined) data alone.
For comparison, the best-fit result of constant $\beta$ case is also shown in the bottom panel.}
\end{figure}

In Fig. \ref{fig1}, using SNe data alone,
we plot the joint $68\%$ and $95\%$ confidence contours for $\{\beta_{0},\beta_{1}\}$ (top panel),
and the $68\%$, $95\%$, and $97\%$ confidence constraints for the evolution of $\beta(z)$ (bottom panel),
for the linear $\beta(z)$ case.
For comparison, we also show the best-fit result of the constant $\beta$ case on the bottom panel.
The top panel shows that $\beta_1 > 0$ at a high CL.
In addition, there is a clear degeneracy between $\beta_0$ and $\beta_1$,
which may be due to the kinematic fact of fitting a linear function.
The bottom panel shows that $\beta(z)$ rapidly increases with $z$.
Moreover, comparing with the best-fit result of constant $\beta$ case,
we can see that $\beta$ deviates from a constant at 6$\sigma$ CL.
It needs to be pointed out that the evolutionary behaviors of $\beta(z)$ depends on the SN samples used.
In \cite{Mohlabeng}, Mohlabeng and Ralston found that, for the Union2.1 SN data, $\beta(z)$ decreases with $z$.
It is of great significance to study why different SN data give different evolutionary behaviors of $\beta(z)$,
and some numerical simulation studies may be required to solve this problem.
We will study this issue in future work.

\begin{itemize}
 \item SN+CMB+GC cases
\end{itemize}

Next, let us discuss the results given by the SN+CMB+GC data.
It should be mentioned that, in order to use the Planck distance priors data,
two new model parameters, $h$ and $\omega_b$, must be added.

In Table \ref{table2}, we make a comparison for the fitting results of constant $\beta$ and linear $\beta(z)$ cases,
where the SN(Combined)+CMB+GC data are used.
Again, we see that adding a parameter of $\beta$ can reduce the values of $\chi_{\rm min}^2$ by $\sim$ 36.
This result is also consistent with the cases of the $\Lambda$CDM, the $w$CDM, and the CPL models \cite{WangNew}.
Therefore, we can conclude that the evolution of $\beta$ is independent of the cosmological models in the background.
This shows that the importance of considering $\beta$'s evolution in the cosmology fits.
In addition, we find that after considering the observational data of CMB and GC,
the parameter ranges of $\Omega_{m0}$, $c$, and $\Omega_{k0}$ become much more reasonable.

\begin{table}
\caption{\textrm{A comparison for the fitting results of constant $\beta$ and linear $\beta(z)$ cases.
The SN(combined)+CMB+GC data are used in the analysis.}}
\label{table2}
\begin{tabular}{|c|c|c|}
  \hline
  \textrm{Parameter} & constant $\beta$ case & linear $\beta(z)$ case \\
  \hline
  $\alpha$ & $1.448^{+0.0760}_{-0.127}$ & $1.416^{+0.097}_{-0.095}$ \\

  $\beta_0$ & $3.270^{+0.082}_{-0.109}$ & $1.403^{+0.359}_{-0.312}$ \\

  $\beta_1$ & \textrm{N/A} & $5.167^{+0.971}_{-0.967}$ \\

  $\Omega_{m0}$ & $0.274^{+0.012}_{-0.016}$ & $0.288^{+0.015}_{-0.013}$ \\

  $h$ & $0.715^{+0.021}_{-0.014}$ & $0.698^{+0.017}_{-0.017}$ \\

  $c$ & $0.687^{+0.057}_{-0.068}$ & $0.768^{+0.112}_{-0.068}$ \\

  $\omega_b$ & $0.02232^{+0.00025}_{-0.00030}$ & $0.02230^{+0.00027}_{-0.00029}$ \\

  $\Omega_{k0}$ & $0.0077^{+0.0039}_{-0.0040}$ & $0.0099^{+0.0051}_{-0.0037}$ \\
    \hline
  $\chi^2_{min}$ & $424.141$ & $388.239$ \\
  \hline
\end{tabular}
\end{table}

We also calculate the best-fit values of ${\cal M}_1$ and ${\cal M}_2$ for the SN+CMB+GC cases.
For the constant $\beta$ case,
we get ${\cal M}_1 = 0.00032$ and ${\cal M}_2 = -0.00251$;
for the linear $\beta(z)$ case,
we obtain ${\cal M}_1 = -0.00053$ and ${\cal M}_2 = -0.00599$.
Again, we can see that ${\cal M}$ has no significant effects on the conclusion of $\beta$'s evolution.

\begin{figure}
\includegraphics[scale=0.25, angle=0]{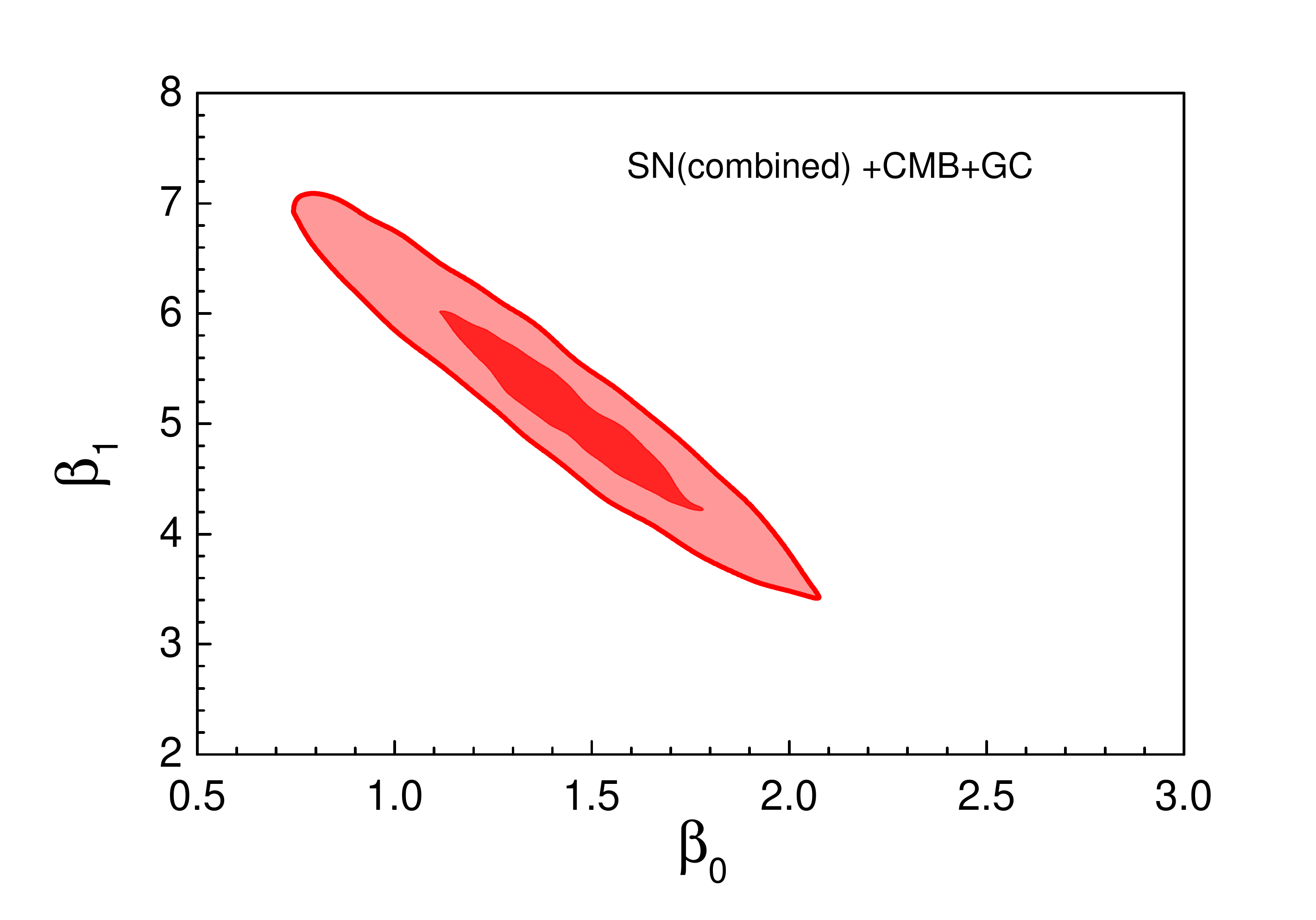}
\includegraphics[scale=0.25, angle=0]{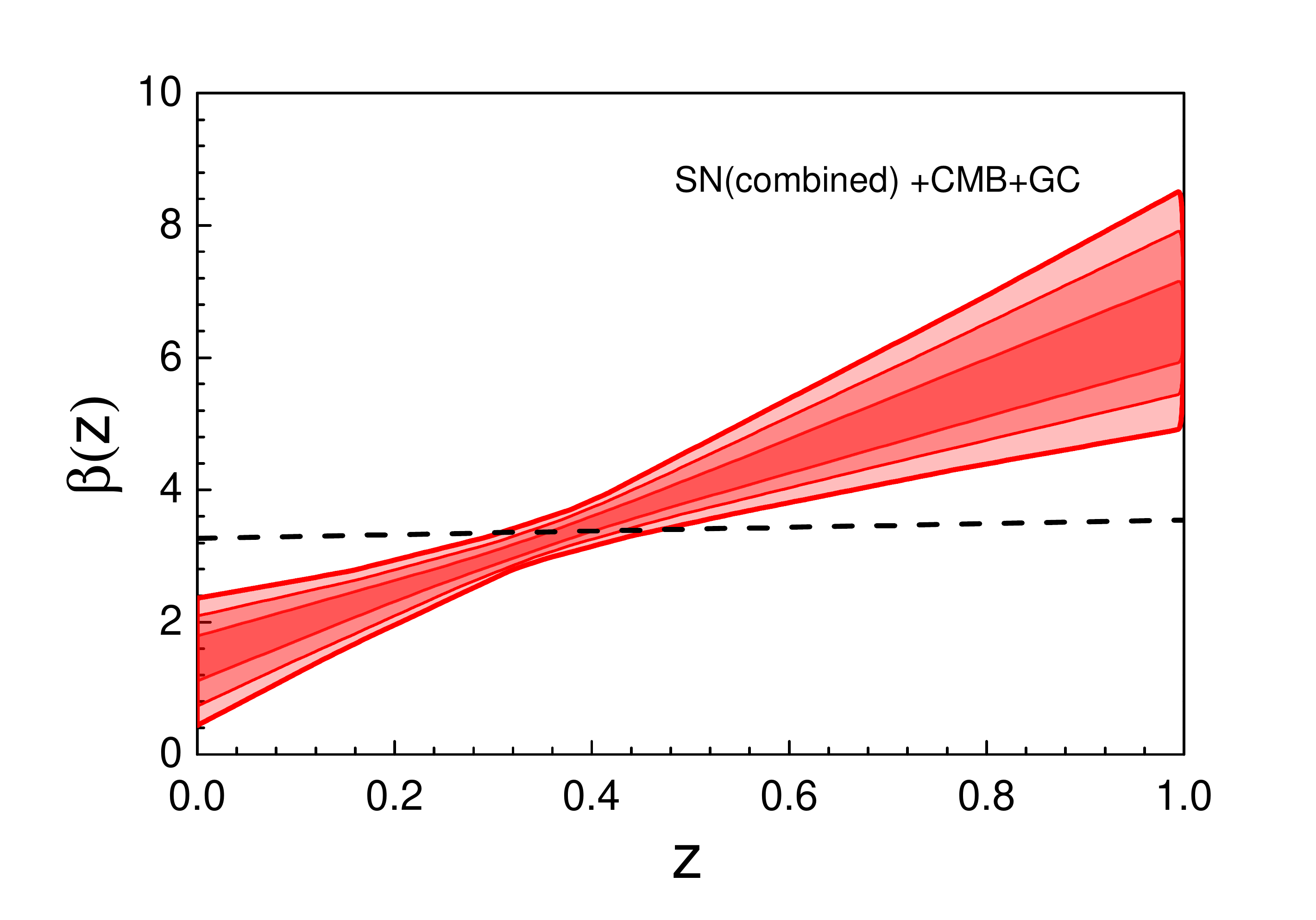}
\caption{\label{fig2}\footnotesize%
The joint $68\%$ and $95\%$ confidence contours for $\{\beta_{0},\beta_{1}\}$ (top panel)
and the $68\%$, $95\%$, and $97\%$ confidence constraints for $\beta(z)$ (bottom panel),
given by the SN(Combined)+CMB+GC data.
For comparison, the best-fit result of constant $\beta$ case is also shown in the bottom panel.}
\end{figure}

Let us discuss the effects of varying $\beta$ with more details.
In Fig.~\ref{fig2}, using SN(combined)+CMB+GC data,
we plot the joint $68\%$ and $95\%$ confidence contours for $\{\beta_{0},\beta_{1}\}$ (top panel),
and the $68\%$, $95\%$, and $97\%$ confidence constraints for the reconstructed evolution of $\beta(z)$ (bottom panel),
for the linear $\beta(z)$ case.
For comparison, we also show the best-fit result of constant $\beta$ case in the bottom panel.
The top panel shows that $\beta_1 > 0$ at a high confidence level,
while the bottom panel shows that $\beta(z)$ rapidly increases with $z$.
In other words, according to this figure,
we find the deviation of $\beta$ from a constant at 6$\sigma$ CL.
This result is consistent with that of Refs. \cite{WangWang} and \cite{WangNew},
and further confirms that the evolution of $\beta$ is insensitive to the DE models
and should be taken into account seriously in the cosmology fits.

\begin{figure}
\includegraphics[scale=0.25, angle=0]{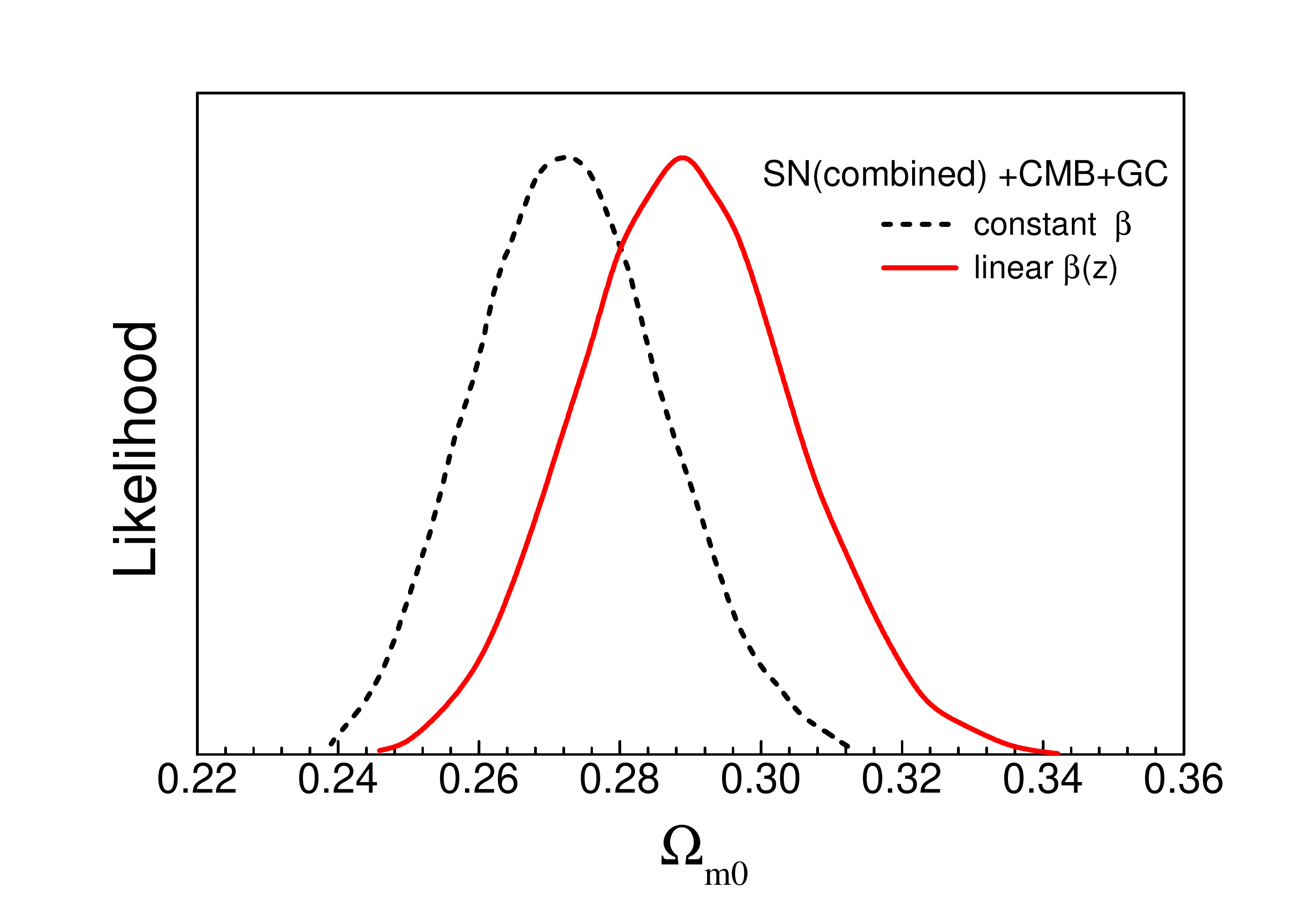}
\includegraphics[scale=0.25, angle=0]{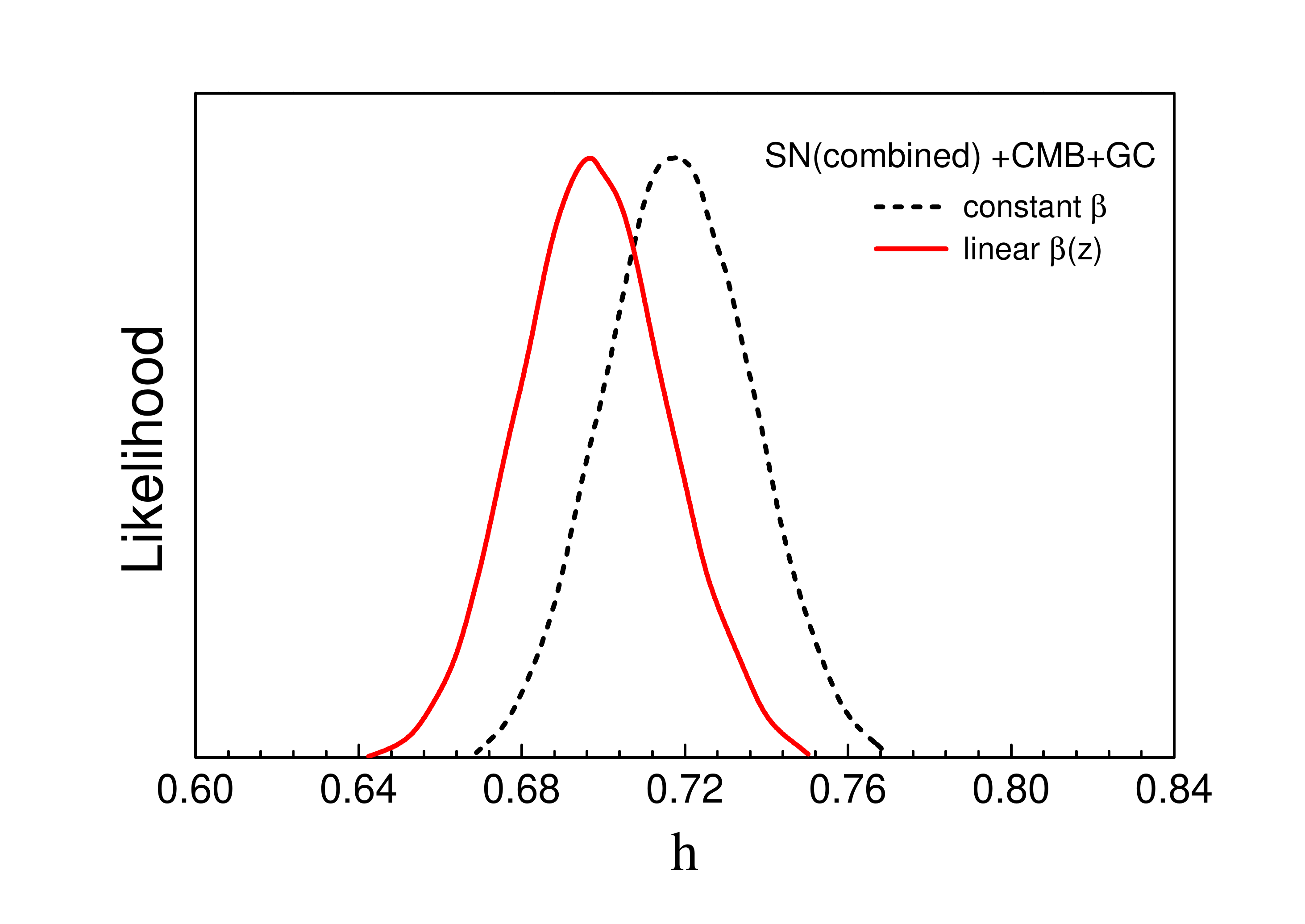}
\includegraphics[scale=0.25, angle=0]{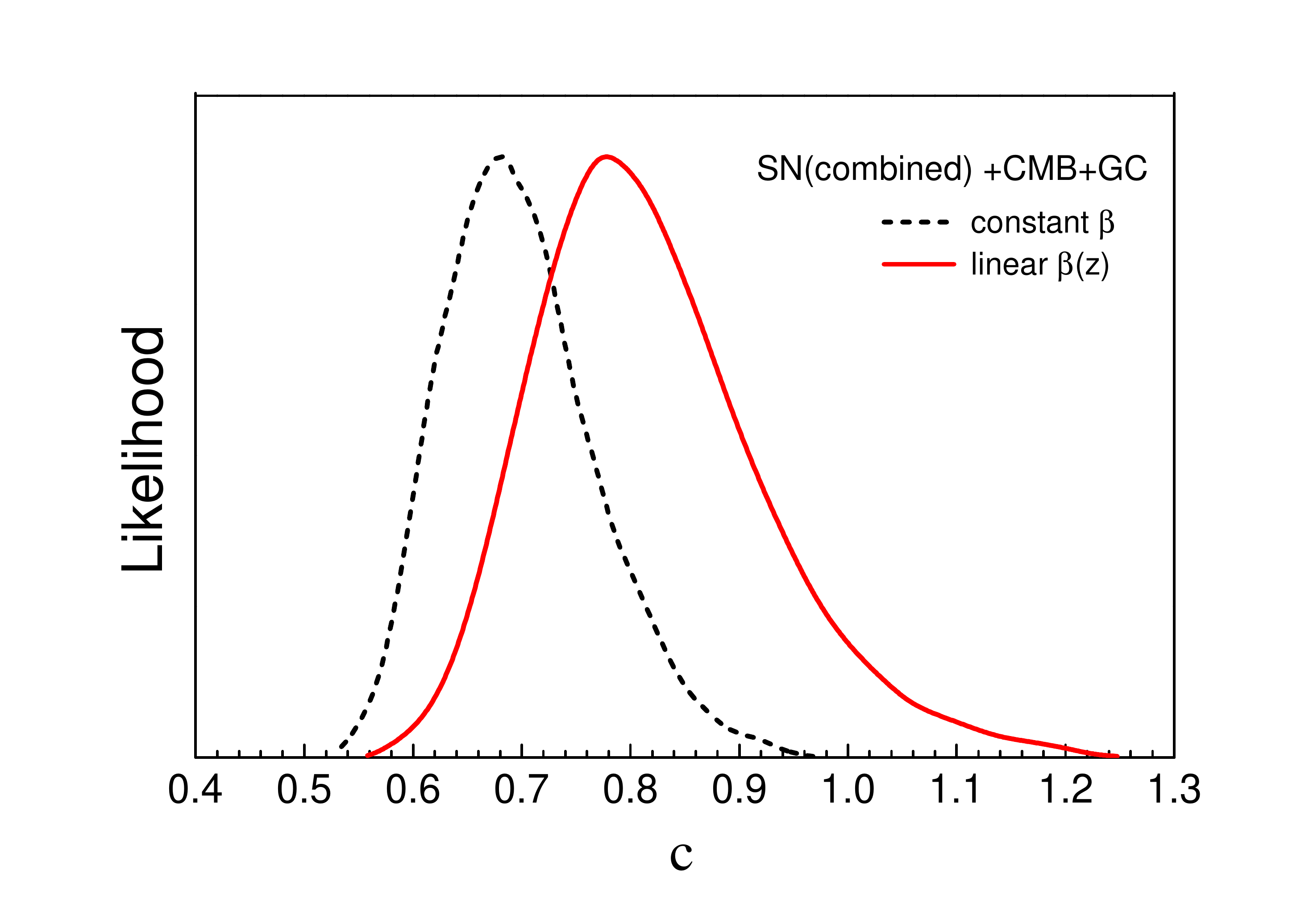}
\caption{\label{fig3}\footnotesize%
The 1D marginalized probability distributions of $\Omega_{m0}$ (top panel), $h$ (central panel), and $c$ (bottom panel),
given by the SN(Combined)+CMB+GC data.
Both the results of constant $\beta$ and linear $\beta(z)$ cases are presented.}
\end{figure}

In Fig.~\ref{fig3}, using the same data,
we plot the 1D marginalized probability distributions of $\Omega_{m0}$, $h$, and $c$,
for both the constant $\beta$ and linear $\beta(z)$ cases.
We find that varying $\beta$ yields a larger $\Omega_{m0}$, a smaller $h$, and a larger $c$.

It must be emphasized that the parameter $c$ plays a very important role in determining the properties of HDE.
For the cases of $c < 1$, $c = 1$, and $c > 1$, the HDE corresponds to a phantom-type, an asymptotic $\Lambda$-type, and a quintessence-type DE, respectively.
The previous studies showed that the best-fit value of this parameter is $c \simeq 0.7-0.8$.
For examples,
in \cite{HDECon1}, using the Gold04+WMAP+LSS data, Zhang and Wu gave $c = 0.81^{+0.23}_{-0.16}$;
in \cite{HDECon2}, using the Constitution+WMAP5+SDSS data, Li et. al. got $c = 0.818^{+0.113}_{-0.097}$;
in \cite{HDECon3}, using the Union2.1+WMAP7+BOSS data, Xu obtained $c = 0.750^{+0.0976}_{-0.0999}$.
But in a later work \cite{PlanckHDE}, Li et. al. found that making use of the Planck data will significantly reduce the value of $c$;
for instance, using the Planck+WP+SNLS3+BAO+HST+lensing data, they obtained $c = 0.563 \pm 0.035$.
In our paper, we find that, adding a parameter of $\beta$ will significantly increase the value of $c$,
and will lead to $c = 0.768^{+0.112}_{-0.068}$, which is consistent with those previous results obtained before the release of Planck data.

\begin{figure}
\includegraphics[scale=0.25, angle=0]{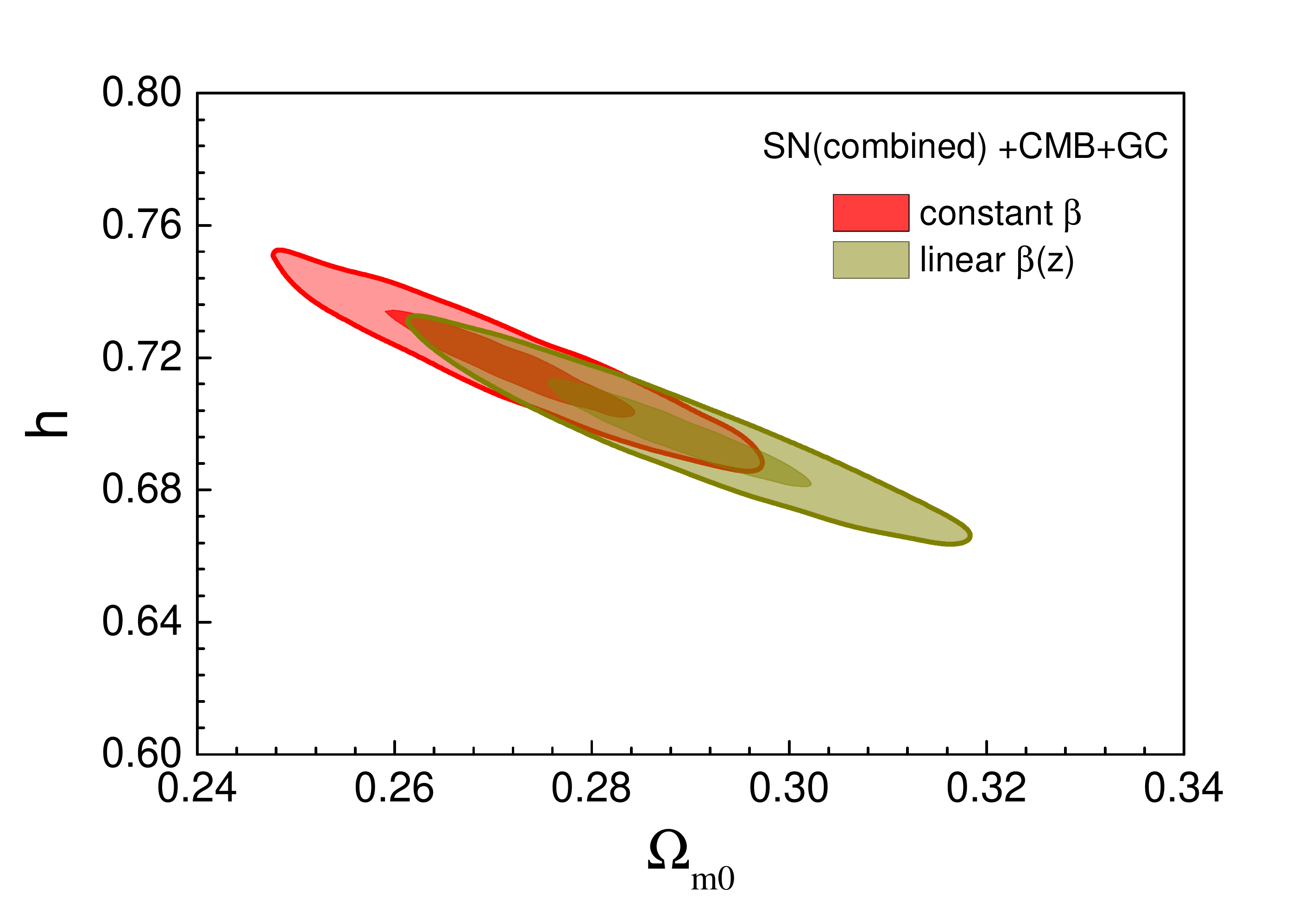}
\includegraphics[scale=0.25, angle=0]{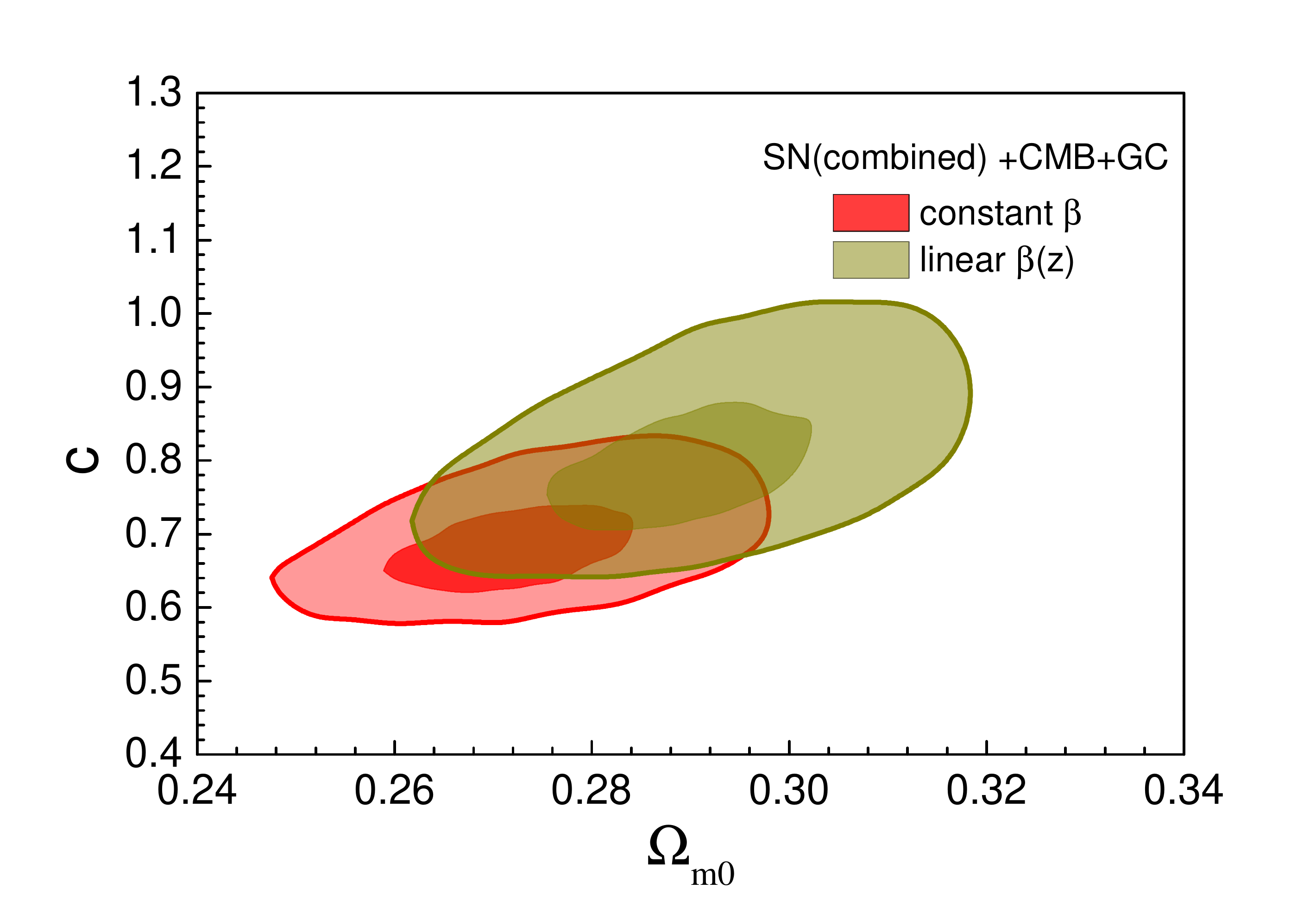}
\includegraphics[scale=0.25, angle=0]{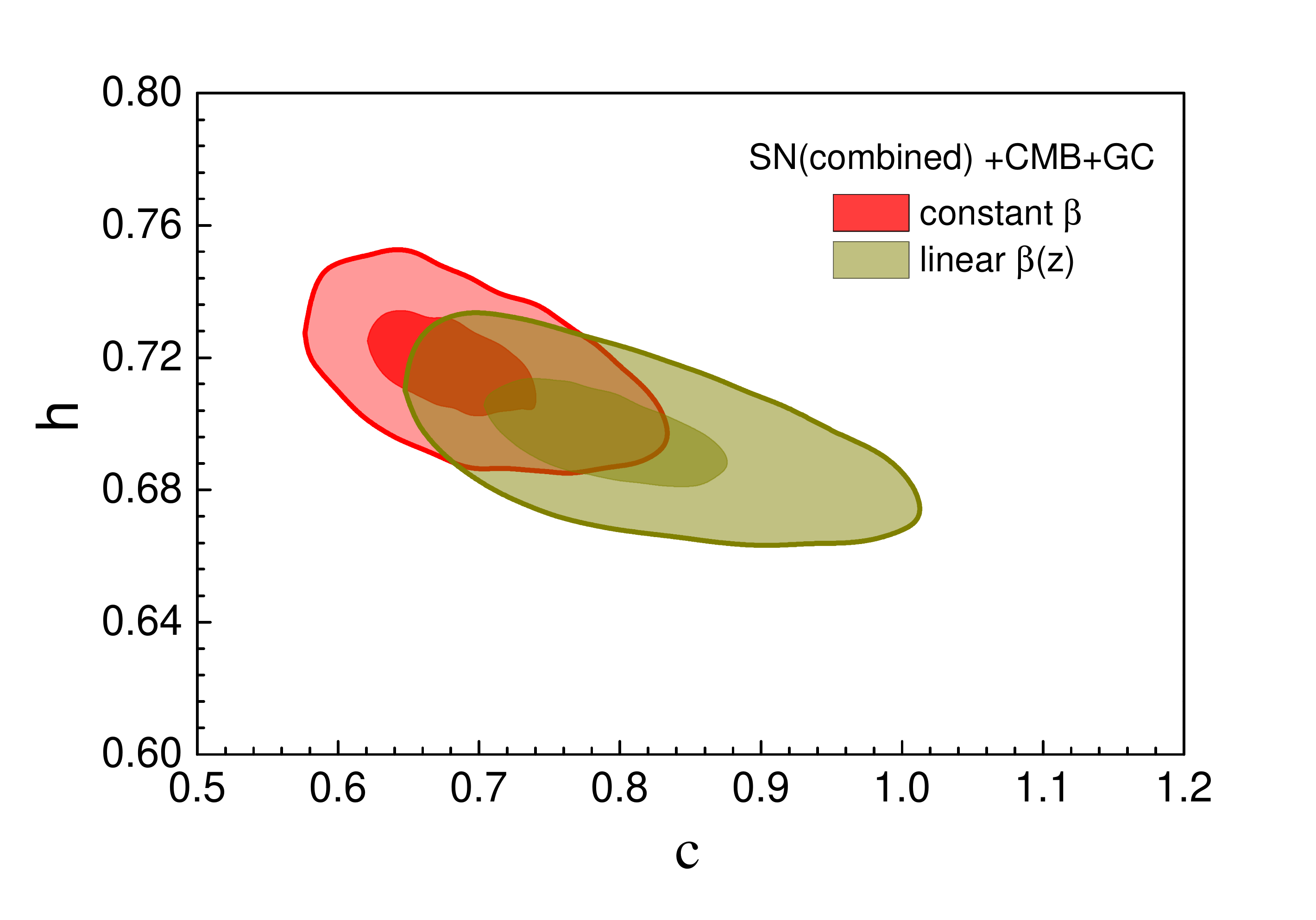}
\caption{\label{fig4}\footnotesize%
The joint $68\%$ and $95\%$ confidence contours
for $\{\Omega_{m0},h\}$ (top panel), $\{\Omega_{m0},c\}$ (central panel), and $\{c,h\}$ (bottom panel),
given by the SN(Combined)+CMB+GC data.
Both the results of constant $\beta$ and linear $\beta(z)$ cases are presented.
}
\end{figure}

In Fig.~\ref{fig4},
we plot the joint $68\%$ and $95\%$ confidence contours
for $\{\Omega_{m0},h\}$, $\{\Omega_{m0},c\}$, and $\{c,h\}$.
Again, we see that varying $\beta$ yields a larger $\Omega_{m0}$, a smaller $h$, and a larger $c$,
compared to the case of assuming a constant $\beta$.
Moreover, we also find that, for these two cases,
the 2$\sigma$ CL ranges of parameter space are quite different.
This means that ignoring the evolution of $\beta$ may cause systematic bias.
In addition, it is clear that $\Omega_{m0}$ and $h$ are strongly anti-correlated;
this is also consistent with the cases of the $\Lambda$CDM, $w$CDM, and CPL models \cite{WangNew}.

\begin{figure}
\psfig{file=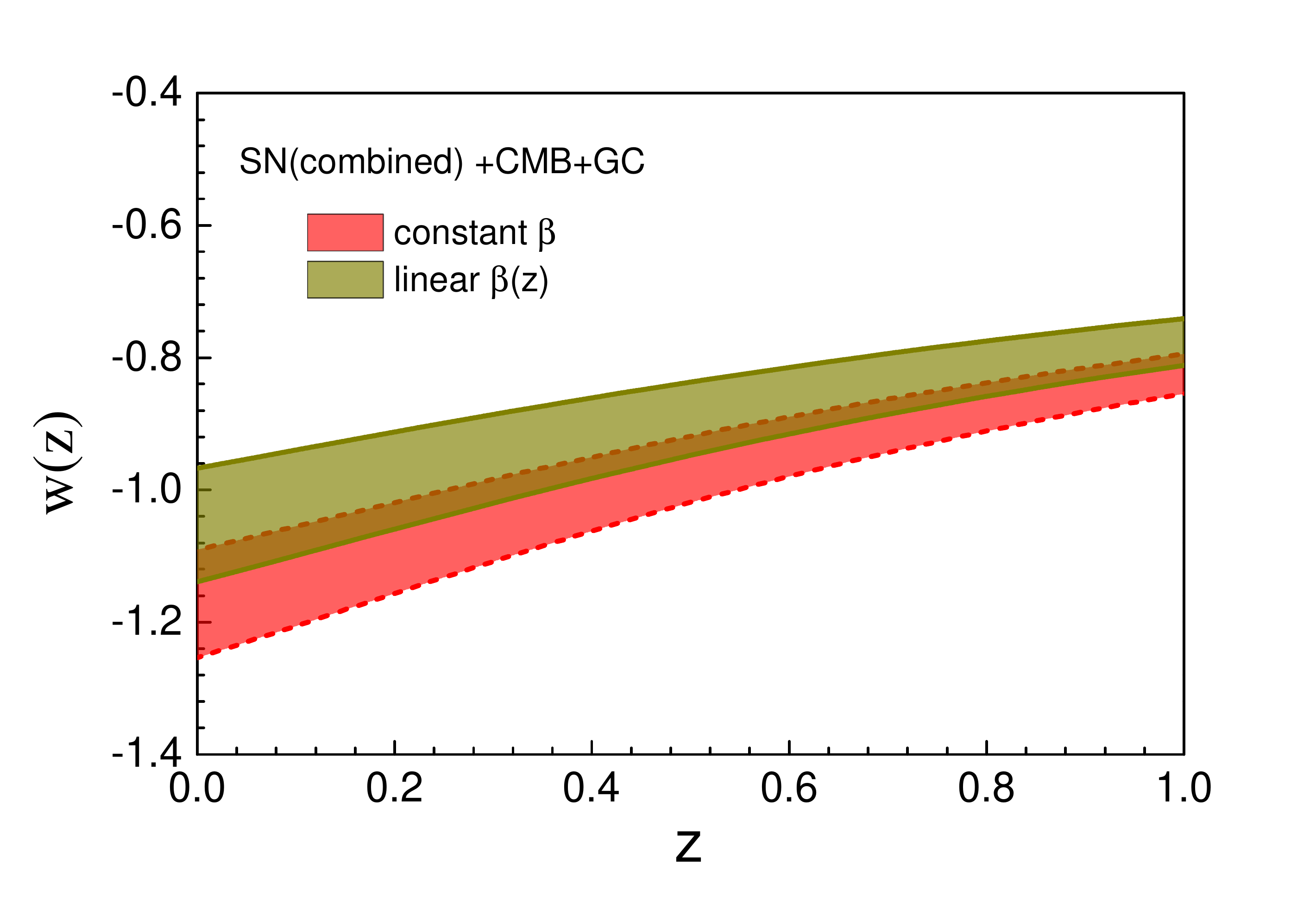,width=3.5in}\\
\caption{\label{fig5}\footnotesize%
The $68\%$ confidence constraints for the EOS $w(z)$ of HDE,
given by the SN(Combined)+CMB+GC data.
Both the results of constant $\beta$ and linear $\beta(z)$ cases are presented.
}
\end{figure}

In Fig.~\ref{fig5},
we plot the $68\%$ confidence constraints for the reconstructed EOS $w(z)$ of HDE.
From this figure,
we see that varying $\beta$ yields a larger $w(z)$:
for the constant $\beta$ case, $w(z=0)<-1$ at 1$\sigma$ CL;
while for the linear $\beta(z)$ case, $w(z=0)$ is still consistent with $-1$ at 1$\sigma$ CL.
Therefore, the results from varying $\beta$ case are in better agreement with a cosmological
constant than those from the constant $\beta$ case.

\subsection{The effects of different light-curve fitters}
\label{sec:difflcf}

In this subsection, we discuss the effects of different light-curve fitters (including ``Combined'', ``SALT2'', and ``SiFTO'').
Notice that we also include the CMB and the GC data.
For simplicity, here we only consider the case of linear $\beta(z)$.
In Table \ref{table3}, we make a comparison for the fitting results given by the ``Combined'', the ``SALT2'', and the ``SiFTO'' SN sets.
An obvious feature of this table is that the differences of various cosmological parameters are very small,
while the differences of SN parameters (including $\alpha$, $\beta_0$ and $\beta_1$) are a little larger.
In the following section, we will discuss this issue with more details.

\begin{table}
\caption{\textrm{A comparison for the fitting results given by the ``Combined'', the ``SALT2'', and the ``SiFTO'' SN sets.
The linear $\beta(z)$ is adopted in the analysis.}}
\label{table3}
\begin{tabular}{|c|c|c|c|}
  \hline
  \textrm{Parameter} & \textrm{Combined} & \textrm{SALT2} & \textrm{SiFTO} \\
  \hline
  $\alpha$ & $1.416^{+0.097}_{-0.095}$ & $1.572^{+0.194}_{-0.151}$ & $1.370^{+0.058}_{-0.081}$ \\

  $\beta_0$ & $1.403^{+0.359}_{-0.312}$ & $1.996^{+0.285}_{-0.249}$ & $1.438^{+0.343}_{-0.359}$ \\

  $\beta_1$ & $5.167^{+0.971}_{-0.967}$ & $3.878^{+0.774}_{-0.835}$ & $5.275^{+0.947}_{-0.894}$ \\

  $\Omega_{m0}$ & $0.288^{+0.015}_{-0.013}$ & $0.285^{+0.017}_{-0.012}$ & $0.284^{+0.018}_{-0.013}$ \\

  $h$ & $0.698^{+0.017}_{-0.017}$ & $0.701^{+0.017}_{-0.018}$ & $0.702^{+0.017}_{-0.019}$ \\

  $c$ & $0.768^{+0.112}_{-0.068}$ & $0.751^{+0.081}_{-0.091}$ & $0.745^{+0.105}_{-0.068}$ \\

  $\omega_b$ & $0.02230^{+0.00027}_{-0.00029}$ & $0.02235^{+0.00023}_{-0.00033}$ & $0.02233^{+0.00024}_{-0.00032}$ \\

  $\Omega_{k0}$ & $0.0099^{+0.0051}_{-0.0037}$ & $0.0101^{+0.0040}_{-0.0043}$ & $0.0091^{+0.0052}_{-0.0037}$ \\
  \hline
\end{tabular}
\end{table}

\begin{figure}
\psfig{file=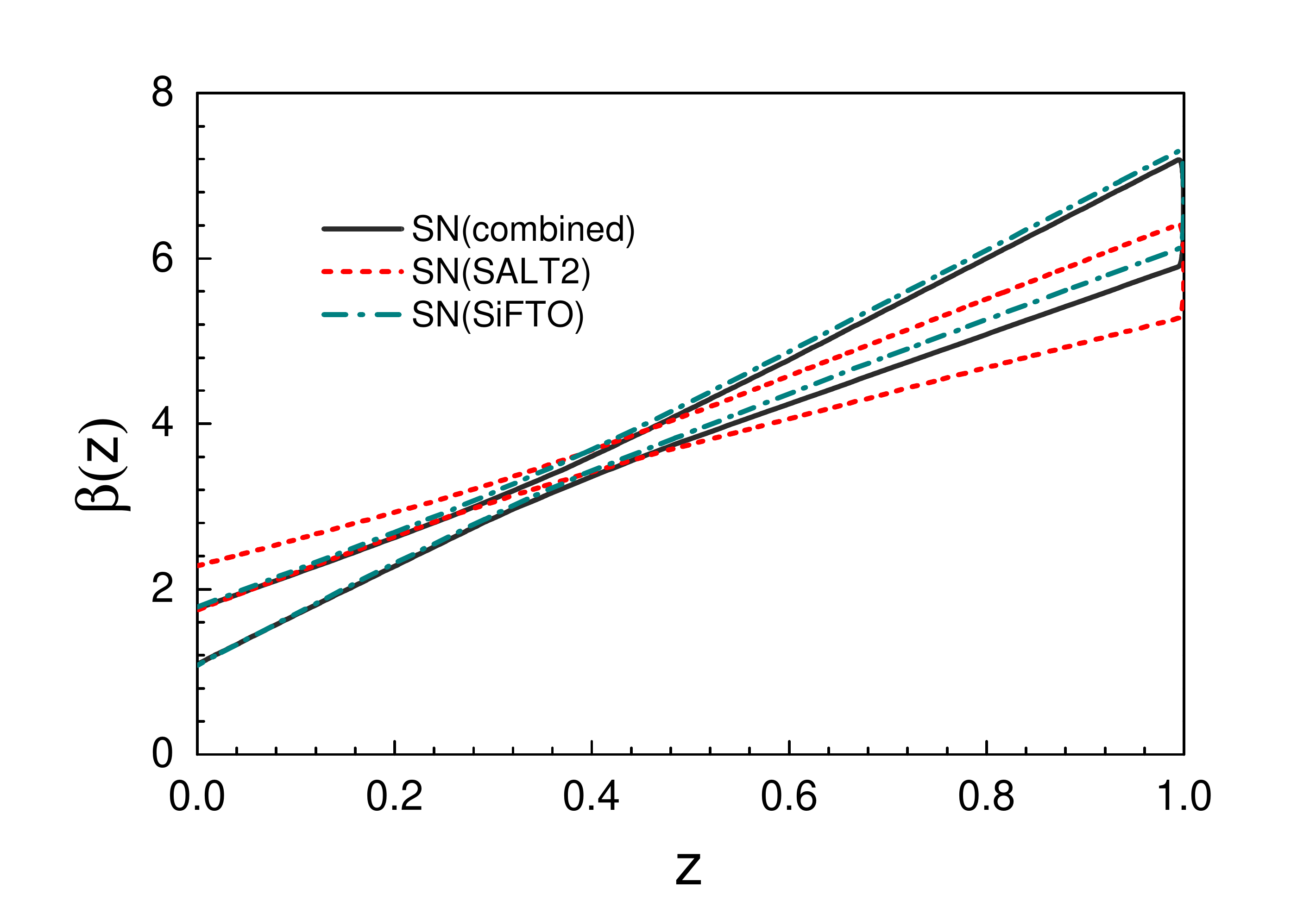,width=3.5in}\\
\caption{\label{fig6}\footnotesize%
The $68\%$ confidence constraints for the evolution of $\beta(z)$ in the HDE model,
given by the ``combined'', the ``SALT2'', and the ``SiFTO'' SN sets.
}
\end{figure}

First, let us focus on the evolution of $\beta$.
In Fig.~\ref{fig6},
we plot the $68\%$ confidence constraints for the reconstructed evolution of  $\beta(z)$ in the HDE model,
given by the ``Combined'', the ``SALT2'', and the ``SiFTO'' SN sets.
It can be seen that the evolution of $\beta$ given by the ``Combined'' set is very close to that given by the ``SiFTO'' SN set;
in contrast, the ``SALT2'' set gives a different $\beta$'s evolution, whose increasing rate is a little smaller.
But for all these three cases, the trends of $\beta(z)$ are still the same,
and all of them deviate from a constant at a high CL.
This result is consistent with the fixed cosmology background case (see Figure 5 of \cite{WangWang}).
Thus, it further confirms that the evolution of $\beta$ is insensitive to the light-curve fitters used to derive the SNLS3 sample.

\begin{figure}
\includegraphics[scale=0.25, angle=0]{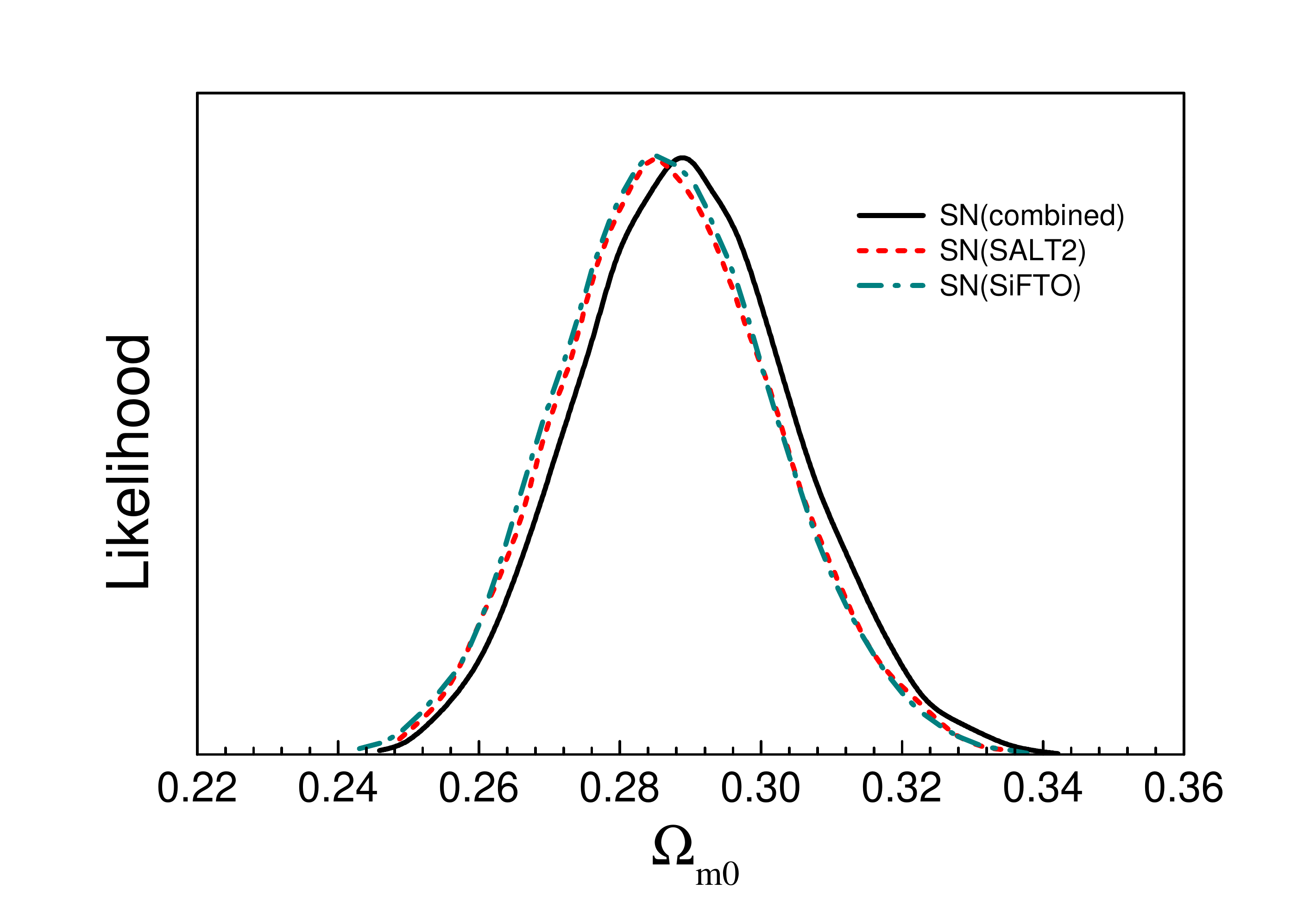}
\includegraphics[scale=0.25, angle=0]{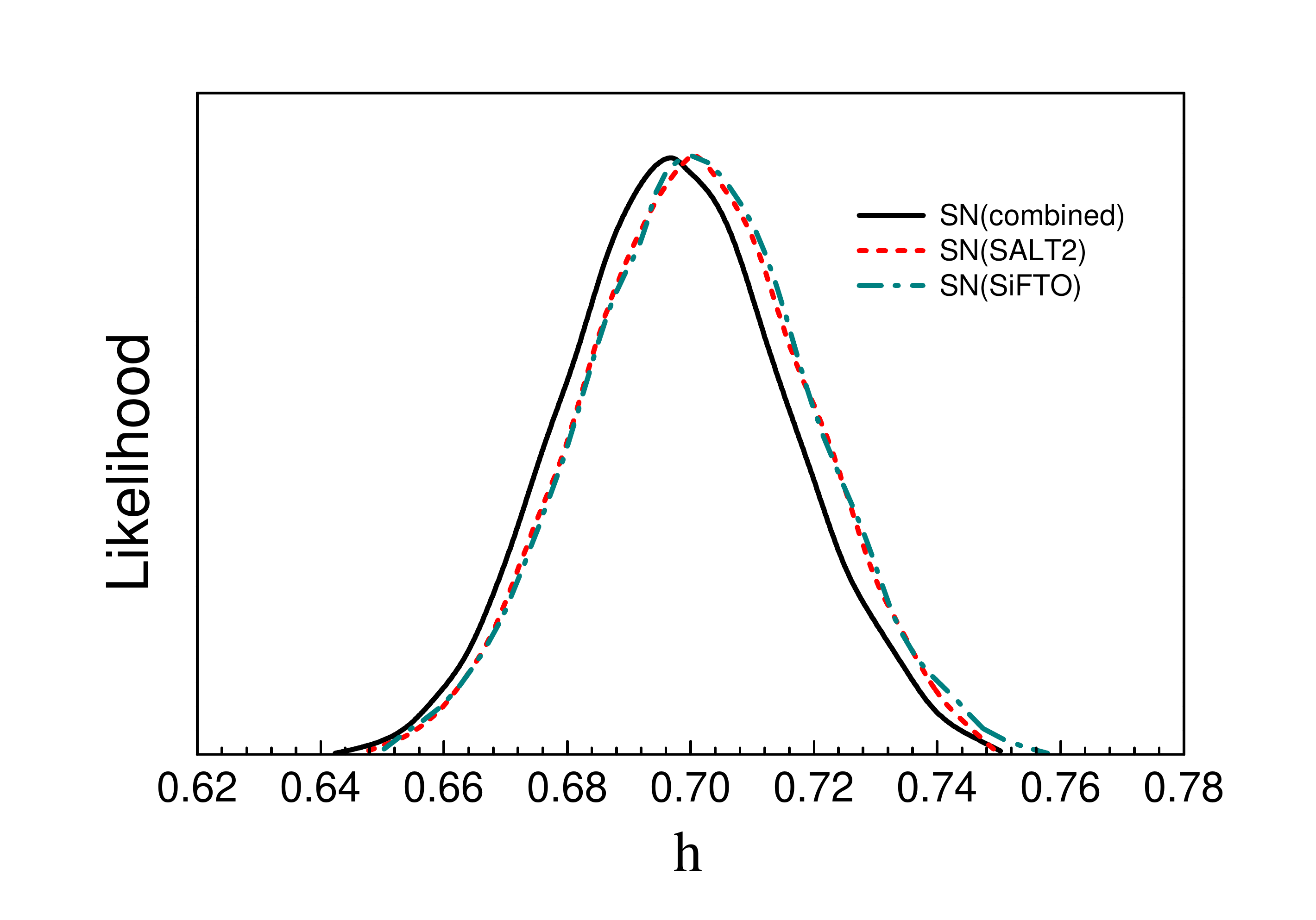}
\includegraphics[scale=0.25, angle=0]{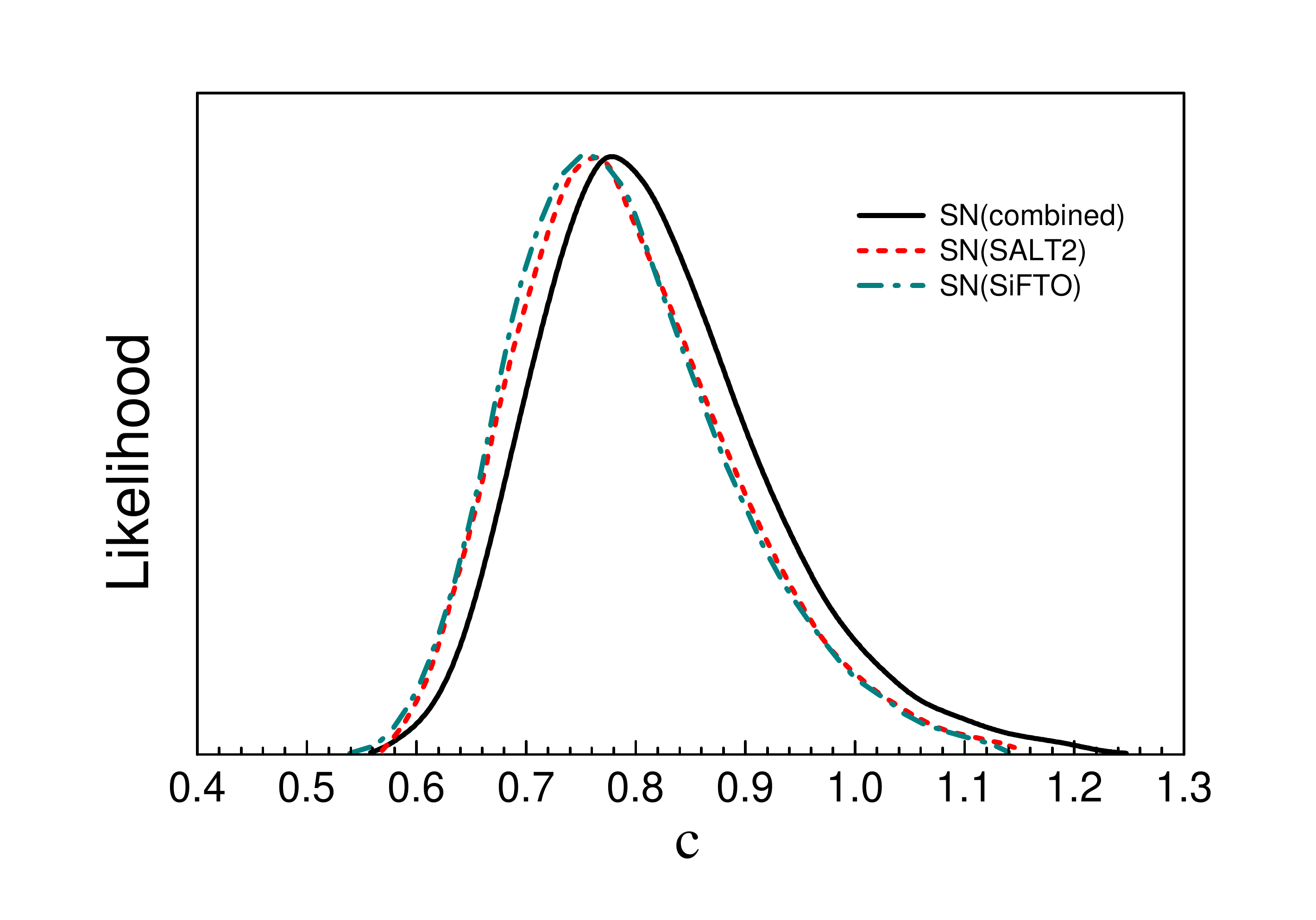}
\caption{\label{fig7}\footnotesize%
The 1D marginalized probability distributions of $\Omega_{m0}$ (top panel), $h$ (central panel), and $c$ (bottom panel),
given by the ``combined'', the ``SALT2'', and the ``SiFTO'' SN sets.}
\end{figure}

Further, let us discuss the effects of different SNLS3 samples on parameter estimation.
In Fig.~\ref{fig7},
we plot the 1D marginalized probability distributions of $\Omega_{m0}$, $h$, and $c$,
given by the ``Combined'', the ``SALT2'', and the ``SiFTO'' SN sets.
It can be seen that the results given by the ``SALT2'' and the ``SiFTO'' SN sets are very close,
while the ``Combined'' set yields a larger $\Omega_{m0}$, a smaller $h$, and a larger $c$.
However, compared to the effects of varying $\beta$ (see Fig.~\ref{fig3}),
the effects of different light-curve fitters are much smaller.

\begin{figure}
\psfig{file=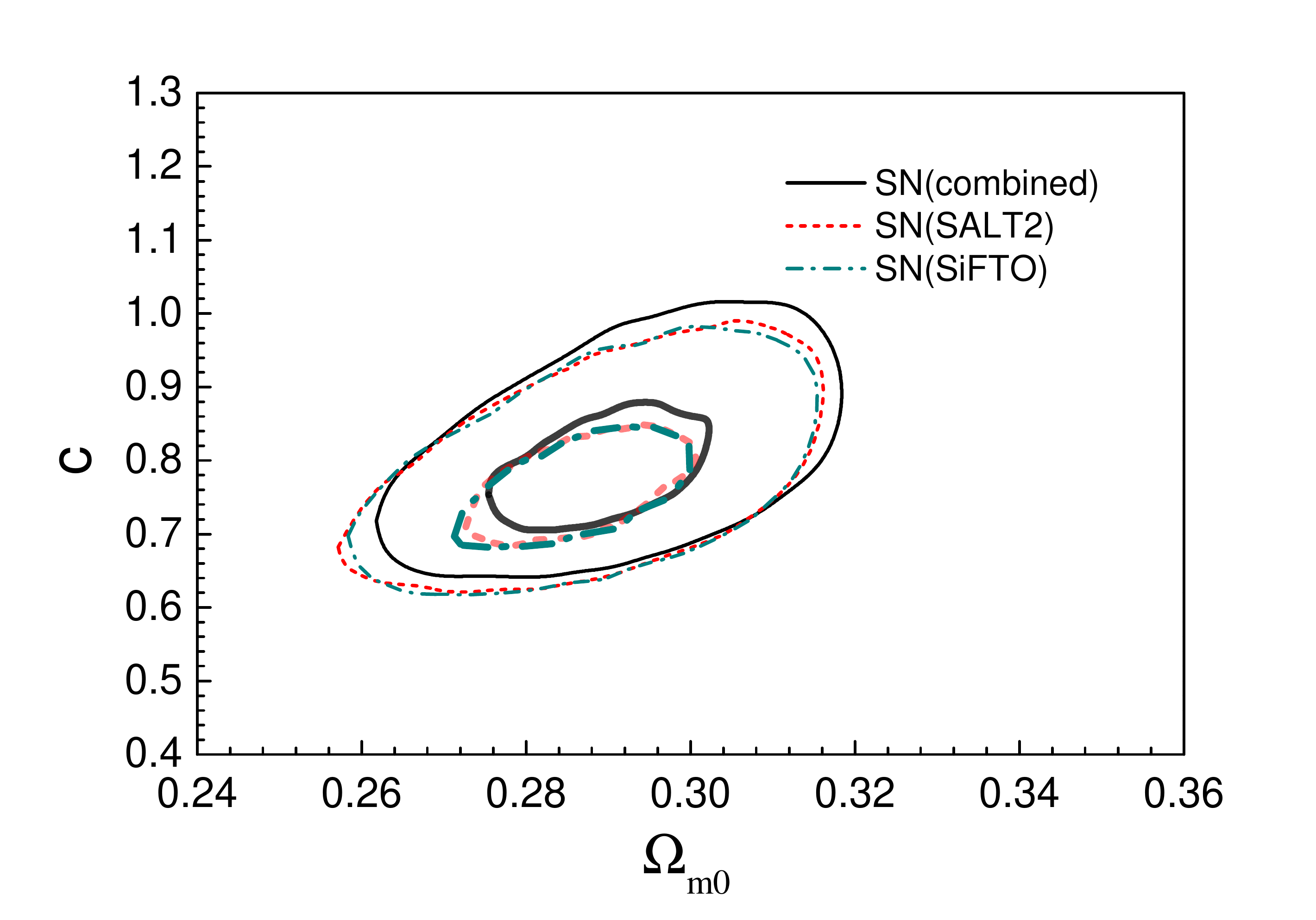,width=3.5in}\\
\caption{\label{fig8}\footnotesize%
The joint $68\%$ and $95\%$ confidence contours for $\{\Omega_{m0},c\}$,
given by the ``combined'', the ``SALT2'', and the ``SiFTO'' SN sets.
}
\end{figure}

In Fig. \ref{fig8},
we plot the joint $68\%$ and $95\%$ confidence contours for $\{\Omega_{m0},c\}$,
given by the three SNLS3 samples.
It is found that the results given by the ``SALT2'' and the ``SiFTO'' SN sets are very close,
while the results given by the ``combined'' set are a little different.
Moreover, compared to Fig.~\ref{fig4},
it can be seen that the effects of different light-curve fitters are much smaller than those of varying $\beta$.

\begin{figure}
\psfig{file=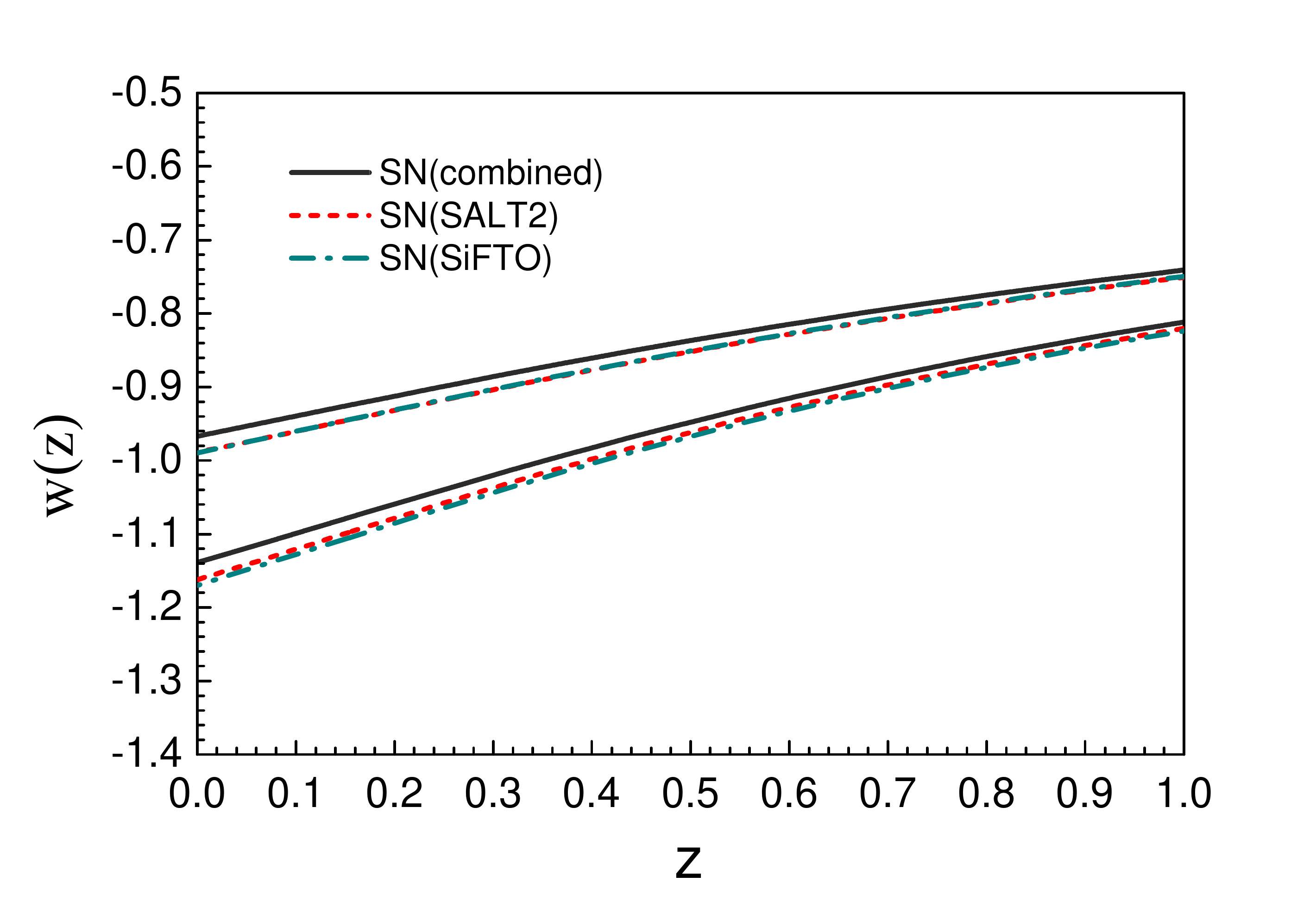,width=3.5in}\\
\caption{\label{fig9}\footnotesize%
The $68\%$ confidence constraints for the EOS $w(z)$ of HDE,
given by the ``combined'', the ``SALT2'', and the ``SiFTO'' SN sets.
}
\end{figure}

At last, we discuss the effects of different light-curve fitters on the EOS $w(z)$.
In Fig.~\ref{fig9},
we make a comparison for the $68\%$ confidence constraints for the EOS $w(z)$ of HDE,
given by the ``Combined'', the ``SALT2'', and the ``SiFTO'' SN sets.
Again, we find that the results given by the ``SALT2'' and the ``SiFTO'' SN sets are very close;
beside, the ``Combined'' set gives a larger $w(z)$.
All these three SN sets yield a $w(z=0)$ that is consistent with $-1$ at 1$\sigma$ CL.
Compared to Fig. \ref{fig5}, again,
we see that the effects of different light-curve fitters are much smaller.

Based on the results of Figs. \ref{fig7}, \ref{fig8}, and  \ref{fig9},
we can conclude that compared to the differences between constant $\beta$ and varying $\beta(z)$ cases,
the effects of different light-curve fitters on parameter estimation are very small.

\section{Discussion and Summary}

It is well known that the
systematic uncertainties of SNe Ia have become the key issue of SN cosmology.
One of the most important systematic uncertainties for SNe Ia is the potential SN evolution,
i.e., the possibility of the evolution of $\alpha$ and $\beta$ with redshift $z$.
In \cite{WangWang},
Wang and Wang found that for the SNLS3 data
there is strong evidence for the evolution of $\beta$.
It must be emphasized that $\beta$'s evolution was not only discovered in the SNLS3 sample,
but also discovered in the other SN datasets (such as Pan-STARRS1 \cite{Scolnic2} and Union2.1 \cite{Mohlabeng}).
Therefore, it is not an isolated phenomenon, and should be taken into account very seriously.

It is clear that the evolution of $\beta$ will have significant effects.
In \cite{WangNew},
using the $\Lambda$CDM model, the $w$CDM model, and the CPL model,
Wang, Li, and Zhang showed that a time-varying $\beta$ has significant impact on parameter estimation;
besides, considering $\beta$'s evolution is rather helpful
for reducing the tension between supernova and other cosmological observations.
To further study the issue of varying $\beta$,
some more specific DE models need to be taken into account.
This is the motivation of this work.

In this paper, we explored the effects of varying $\beta$ on the cosmological constraints of HDE model.
In addition to the SNLS3 data,
we have also used the latest Planck distance priors data \cite{WangWangCMB},
as well as the latest GC data extracted from SDSS DR7 \cite{ChuangWang12} and BOSS \cite{Chuang13}.
In addition, we have also studied the effects of different light-curve fitters on parameter estimation.

We found that for both the SNe alone and the SN+CMB+GC cases,
adding a parameter of $\beta$ can reduce the best-fit values of $\chi^2$ of HDE model by $\sim$ 36
(see Table~\ref{table1} and  Table~\ref{table2});
it means that $\beta$ deviates from a constant at the 6$\sigma$ CL (see Figs. \ref{fig1} and \ref{fig2}).
This result is consistent with those of the $\Lambda$CDM, the $w$CDM, and the CPL models.
This implies that the evolution of $\beta$ is insensitive to the DE models in the background
and should be taken into account seriously in the cosmology fits.

Adopting SN+CMB+GC data,
we found that compared to the constant $\beta$ case,
varying $\beta$ yields a larger $\Omega_{m0}$ and a smaller $h$ (see Figs. \ref{fig3});
moreover, varying $\beta$ will significantly increase the value of $c$,
consistent with the constraint results obtained before the Planck data.
In addition, for these two cases,
the 2$\sigma$ CL ranges of parameter space are quite different (see Figs. \ref{fig4}).
This indicates that ignoring the evolution of $\beta$ may causes systematic bias.

Varying $\beta$ also yields a larger $w(z)$:
for the constant $\beta$ case, $w(z=0)<-1$ at 1$\sigma$ CL;
while for the linear $\beta(z)$ case, $w(z=0)$ is consistent with $-1$ at 1$\sigma$ CL (see Fig.~\ref{fig5}).
So, the results from the varying $\beta$ SN data are in better agreement with a cosmological
constant than those from the constant $\beta$ SN data.

We found that the evolution of $\beta$ given by the ``combined'' and the ``SiFTO'' SN sets are very close;
while the ``SALT2'' set will give a different $\beta$'s evolution, whose increasing rate is a little smaller (see Fig. \ref{fig6}).
We also found that the cosmology-fits results given by the ``SALT2'' and the ``SiFTO'' SN sets are very close,
while the ``Combined'' set yields a larger $\Omega_{m0}$, a smaller $h$, a larger $c$, and a larger $w(z)$
(see Figs. \ref{fig7}, \ref{fig8} and \ref{fig9}).
However, compared to the differences between constant $\beta$ and time-varying $\beta(z)$ cases,
the effects of different light-curve fitters are very small.

In this paper, only the potential SN evolution is taken into account.
Some other factors, such as the evolution of $\sigma_{int}$ \cite{Kim2011},
may also cause systematic uncertainties for SNe Ia.
This issue deserves further study in the future.


\begin{acknowledgments}
We thank Yun-He Li for the helpful discussions.
We are also grateful to Alex Conley for providing us with the SNLS3 covariance matrices that allow redshift-dependent $\alpha$ and $\beta$.
We acknowledge the use of CosmoMC.
SW is supported by the National Natural Science Foundation of China under Grant No.~11405024
and the Fundamental Research Funds for the Central Universities under Grant No.~N130305007.
XZ is supported by the National Natural Science Foundation of China under Grant No.~11175042
and the Fundamental Research Funds for the Central Universities under Grant No.~N120505003.
\end{acknowledgments}

\end{document}